\documentclass[10pt,journal,compsoc]{IEEEtran}
\usepackage{xcolor}
\usepackage{cite}
\usepackage{amsmath,amssymb,amsfonts}
\usepackage{algorithmic}
\usepackage{graphicx}
\usepackage{textcomp}
\usepackage{wrapfig}
\usepackage{diagbox}
\usepackage{mathtools}
\usepackage{textcomp}
\usepackage{tikz}
\usepackage{svg}
\svgsetup{
    inkscapepath=./
}
\usepackage{algorithm2e}
\usepackage{multirow}
\usepackage{tabularx}
\usepackage{booktabs}
\usepackage{balance}
\usepackage{array}
\newcolumntype{L}[1]{>{\raggedright\arraybackslash}p{#1}}

\def\gap{1ex}
\abovedisplayskip\gap
\belowdisplayskip\gap
\abovedisplayshortskip\gap
\belowdisplayshortskip\gap

\newcommand{\abovetable}{0.1cm}
\newcommand{\belowtable}{-0.4cm}
\newcommand{\betweentablecaption}{-0.2cm}

\newcommand{\belowfig}{-0.4cm}
\newcommand{\belowfigcaption}{-0.3cm}

\usetikzlibrary{shapes.geometric, arrows, positioning, fit, chains, backgrounds}
\tikzstyle{flowchartnode} = [rectangle, rounded corners, 
minimum width=1.5cm, 
minimum height=1cm,
text centered, 
text width=2cm, 
draw=black, 
fill=white,
font=\small,
]

\definecolor{darkgreen}{rgb}{0.0, 0.5, 0.0} 
\definecolor{darkblue}{rgb}{0.0, 0.0, 0.5} 

\RestyleAlgo{ruled}
\SetKwComment{Comment}{/* }{ */}
\SetKwInOut{Input}{Input}
\SetKwInOut{Output}{Output}

\newcommand{\doubleceil}[1]{\left\lceil\!\!\left\lceil #1 \right\rceil\!\!\right\rceil}

\begin{document}

\begingroup
\thispagestyle{empty}
\onecolumn
{\Huge\bfseries \noindent IEEE Copyright Notice\par}

\vspace{1cm}
\noindent© 2026 IEEE.  Personal use of this material is permitted.  Permission from IEEE must be obtained for all other uses, in any current or future media, including reprinting/republishing this material for advertising or promotional purposes, creating new collective works, for resale or redistribution to servers or lists, or reuse of any copyrighted component of this work in other works.

\newpage
\endgroup

\clearpage
\setcounter{page}{1}
\twocolumn

\title{
    WiRM: Wireless Respiration Monitoring Using Conjugate Multiple Channel State
    Information and Fast Iterative Filtering in Wi-Fi Systems
}

\author{James Rhodes, Lawrence Ong, and Duy T. Ngo
    \thanks{
        This article has been accepted for publication in IEEE Transactions on
        Mobile Computing. This is the authors’ version. The final published
        version will be available in IEEE Xplore. DOI:
        10.1109/TMC.2026.3711502
    }
    \thanks{
        James Rhodes, Lawrence Ong, and Duy T. Ngo are with the School of Engineering,
        The University of Newcastle, Australia (e-mail: j.rhodes@uon.edu.au, lawrence.ong@newcastle.edu.au, duy.ngo@newcastle.edu.au).
    }
     \thanks{This work is supported in part by The University of Newcastle through an Australian Government Research Training Program Scholarship (Strategic Engagement Scheme).}
}

\markboth{IEEE TRANSACTIONS ON MOBILE COMPUTING, VOL. XX, NO. X, MONTH XX, 2026}%
{}

\maketitle

\begin{abstract}
Monitoring respiratory health with the use of channel state information (CSI)
has shown promising results. Many existing methods focus on monitoring only the
respiratory rate, while others focus on monitoring the motion of the chest as a
patient breathes, which is referred to as the \textit{respiratory waveform}.
This paper presents WiRM, a two-staged approach to contactless respiration
monitoring. In the first stage, WiRM improves upon existing respiratory rate
estimation techniques by using conjugate multiplication for phase sanitisation
and the adaptive multi-trace carving (AMTC) algorithm for tracing how the
respiratory rate changes over time. When compared against four
state-of-the-art methods, WiRM has achieved an average reduction of
$34.7\%$ in respiratory rate root mean squared error (RMSE). In the second
stage, WiRM uses this improved respiratory rate estimate to inform the
decomposition and selection of the respiratory waveform from the CSI data. WiRM
delivers a $7.9\%$ improvement in average absolute correlation with the
ground truth respiratory waveform. Within the literature, it is difficult to
compare the robustness of existing algorithms in noisy environments. In this
paper, we develop a purpose-built simulation toolkit to evaluate the robustness
of respiration monitoring solutions under ambient motion interference and
various noise conditions, including thermal, multiplicative, and phase noise.
Our results show that WiRM demonstrates improved or comparable resilience to
these common noise sources and interference scenario. 
\end{abstract}


\section{Introduction} \label{sec:introduction}

\IEEEPARstart{R}{espiration} monitoring is a key diagnostic tool utilised by
medical practitioners to monitor a patient's health. The monitoring of
respiratory health can include a patient's respiratory rate or the motion of the
patient's chest in a direction normal to the torso as they breathe, referred to as the \textit{respiratory waveform}. The accurate tracking of both the respiratory rate and the respiratory waveform can aid in the diagnosis
of many illnesses.

The gold standard methodology for monitoring respiratory rate in hospitals is
for a healthcare worker to manually count the number of inhalations over one-minute periods \cite{ref:resp_rate_inaccuracy}. This is time-consuming for patients and professionals, introducing human error into the monitoring. The
state-of-the-art automatic respiratory rate monitoring, e.g., during
a polysomnography test, requires patients to wear several uncomfortable
pieces of equipment such as a belt to monitor the motion of the chest and/or
pressure transducers to monitor the breath. This can cause discomfort to the
patients, potentially skewing the results of sleep tests. 

Similar to the measurement of respiratory rate, the respiratory waveform is commonly measured manually or via uncomfortable
equipment. Unlike respiratory rate which provides a single scalar value, the respiratory waveform captures
the full temporal dynamics of breathing. This richer signal enables a broader
range of diagnostic applications, including monitoring airflow and volume
changes, identifying apnoea, and identifying abnormal chest wall movements
\cite{ref:uses_of_breath_waveform}. These capabilities make the waveform more
informative than respiratory rate alone, the latter of which can often be readily derived from the
waveform.

While optical methods for contactless respiration monitoring aim to solve the
discomfort problem, the need for constant visual surveillance of a
patient incurs privacy concerns. The need for a respiratory rate monitoring solution that is cost-effective, privacy-preserving, contactless, and automated is therefore apparent.

Wireless technologies provide an opportunity for contactless sensing of motion
and, by not using cameras, offer a more private experience for patients.
Wireless respiratory rate monitoring is an active area of research with
advancements utilising acoustics\cite{ref:acoustic_one, ref:acoustic_two}, Radio
Frequency Identification (RFID)\cite{ref:rfid_one, ref:rfid_two}, radar
\cite{ref:radar_one, ref:deep_breath}, and commodity Wi-Fi \cite{ref:wifi_one,
ref:wifi_two}. Except for Wi-Fi, all of these methods require specialised
equipment such as microphones, RFID tags, or costly radar systems. This makes
Wi-Fi-based approaches stand out thanks to the low cost and proliferation of
Wi-Fi systems in homes and public spaces. 

Direct respiratory rate estimation continues to struggle in environments with a
low signal-to-noise ratio and does not provide medical practitioners with a
respiratory waveform for diagnostic purposes \cite{ref:smars, ref:tr_breath}.
Methods that compute the waveform directly are generally more sensitive to
noise. They also employ decomposition techniques that require an additional,
usually heuristic-based, method to extract the breath signal from the set of
decomposed signals \cite{ref:pos_free_breath}. Furthermore, since there is a
severe lack of datasets or simulation methods in this area of research,
different sources of data are being used, making it challenging to compare the
performance of different methods.

This paper proposes a novel Wireless Respiration Monitoring (WiRM) solution to
address these issues. Specifically:
\begin{itemize}
    \item{
        \textbf{WiRM -- Stage 1}: We first propose a new respiratory rate
        estimation algorithm that utilises adaptive multi-trace carving with an
        empirically determined respiratory rate transition probability function.
        The aim is to improve respiratory rate tracking performance. Our
        proposed rate estimation algorithm demonstrates an improvement of $34.7\%$
        in average Root Mean Square Error (RMSE) compared to the
        state-of-the-art techniques.
    }
    \item{
        \textbf{WiRM -- Stage 2}: We then propose a new method for estimating
        respiratory waveforms from previous respiratory rate estimations,
        utilising the fast iterative filtering algorithm. Our proposed waveform
        estimation method delivers an improvement of $7.9\%$ in average
        absolute correlation over the state-of-the-art techniques. 
    } 
    \item{
        \textbf{Simulation Toolkit for Performance Evaluation in Noisy
        Environments}: We develop a purpose-built simulation toolkit to
        systematically evaluate the performance of our proposed WiRM algorithm
        against ideal CSI models under three common noise sources: thermal,
        multiplicative, and phase. Numerical results with practical parameter
        settings show that WiRM demonstrates substantial robustness across all
        three noise types, consistently matching or exceeding the performance of
        existing leading algorithms. We further evaluate the performance of WiRM
        in a simulated interference scenario where WiRM shows comparable
        performance to the current state-of-the-art.
    }
\end{itemize}

\section{Related Works}
\label{sec:related_works}

The first advances in Wi-Fi-based sensing use received signal strength as an
indicator of various movements within the environment \cite{ref:rss}. The use of
signal power was not fine-grained enough for more advanced use cases of sensing.
To overcome this granularity problem, researchers turn to utilising the
Wi-Fi's channel state information (CSI).

Some key application domains enabled by CSI's fine-grained
measurements include fall-detection\cite{ref:wi_fall}, human activity
recognition (HAR) \cite{ref:sharp}, gait estimation \cite{ref:gaitway}, speed
estimation \cite{ref:wi_speed}, indoor localisation and tracking
\cite{ref:localization_self_supervised}, and vital sign monitoring
\cite{ref:wital}. All of these research areas aim to address a common problem of detecting minute changes in CSI measurements relating to direct physical movement in the environment.

Changes in the environment are reflected in both the phase and magnitude of the
measured CSI. As the phase is often affected by noise, pre-processing or estimation techniques are needed to remove it. The three broad categories of methods for detecting changes in the environment using CSI measurements
include using the magnitude of CSI (or similarly, CSI energy)
\cite{ref:wi_speed,ref:smars,ref:wi_resp}, the phase of CSI
\cite{ref:phaser,ref:tensorbeat,ref:phasebeat,ref:fullbreathe,ref:multisense,ref:sharp},
or a combination of both \cite{ref:pos_free_breath, ref:wirim,
ref:concentric_circle} to avoid ``blind-spots''
\cite{ref:human_resp_detection_with_commodity_wifi}.

\subsection{Direct Respiratory Rate Estimation}
\label{ssec:direct_resp_rate_estimation_lit_review}

TR-BREATH is one of the first methods to realise simultaneous, multi-user
respiration rate detection. Specifically, it utilises the spatial and temporal focusing effects of the
time reversal principle (TR) and time reversal resonating strength (TRRS) to
detect minute changes in the environment, allowing for the calculation of the
respiratory rate \cite{ref:tr_breath}. The TRRS is analogous to a correlation
coefficient and accurately describes how the environment changes with time. The
TRRS is computed across a time window, followed by applying the
Root Multiple Signal Classification (Root-MUSIC) algorithm to gather
candidate breathing rates. The output of the Root-MUSIC algorithm is used to
compute statistics about the breathing data, which is then fed into a support
vector machine to detect whether breath is present or not. The output of the
Root-MUSIC algorithm is also used in conjunction with affinity propagation
clustering and likelihood assignment to determine the likely breathing rates
\cite{ref:tr_breath}. As the Root-MUSIC algorithm incurs high computational
complexity, TR-BREATH is more suitable for off-line computations. 

A competing approach to detecting motion within CSI uses the autocorrelation
function (ACF). This approach has motion detection capabilities similar to the
TR principle \cite{ref:tr_acf_connection} and was first applied to respiration
monitoring by \textit{SMARS: Sleep Monitoring via Ambient Radio Signals}
\cite{ref:smars}. The SMARS method measures the period of breathing based on the
distance between the peaks and troughs of the ACF of the CSI energy to estimate
the breathing rate.

Building on SMARS, Wang \emph{et al.} compute an ACF at each time step to construct an
\textit{ACF spectrum} in their study, \textit{WiResP: A Robust Wi-Fi-Based
Respiration Monitoring via Spectrum Enhancement} \cite{ref:wi_resp}. Image-processing techniques
are then applied to the \textit{ACF spectrum}, which is subsequently compared to
a template \textit{ACF spectrum} known to contain breathing patterns, enhancing
breath detection capabilities.

It should be noted that both WiResP and SMARS use only the CSI energy, which has been shown to create
``blind-spots'' within a given room. Due to the phase noise, using both the phase and the magnitude of
the CSI does not necessarily solve the blind-spot problem \cite{ref:fullbreathe}. In this paper, we use conjugate multiplication to cancel out phase noise, enabling the use of both the magnitude and phase of the CSI to address the
blind-spot problem \cite{ref:fullbreathe}. We then apply the ACF to detect the
periodic respiratory rate within CSI data as described in
Section~\ref{sec:proposed_algorithm}. Independent and concurrent to our
research, Wang \emph{et al.} \cite{ref:acc_est_of_path_length_changes} have used
conjugate multiplication and the parametric symmetric ACF in a different application, i.e., fall detection.  

An emerging area of research is the application of CSI-based sensing techniques
to a more easily accessible data source, the beamforming feedback information.
This compressed representation of CSI is transmitted as plain, unencrypted text,
allowing devices such as packet sniffers to access this information for
sensing purposes. One example is the work of Kanda \emph{et al.}
\cite{ref:bff_sensing}, where digital signal processing techniques are applied to
recover the respiratory rate and heart rate of a stationary person in a room.

\subsection{Respiratory Waveform Estimation} 
\label{ssec:resp_waveform_lit_review}

Methods for extracting the respiratory waveform generally share
similar steps. They include pre-processing (such as filtering or smoothing
the raw CSI data), signal selection (such as choosing between CSI
amplitude, phase, or other metrics), and a decomposition technique to extract
potential breath waveforms, from which the final breath waveform can be
determined. Popular decomposition methods include empirical mode decomposition (EMD) \cite{ref:emd_one},
independent component analysis \cite{ref:deep_breath,ref:ica_one,
ref:multisense}, canonical polyadic decomposition \cite{ref:tensorbeat},
discrete wavelet transform \cite{ref:phasebeat}, variational mode decomposition
(VMD) \cite{ref:vmd_one,ref:pos_free_breath}, and principal component analysis
(PCA) \cite{ref:pos_free_breath}.

TensorBeat is among the first to realise respiratory-waveform estimation from CSI data
\cite{ref:tensorbeat}. It addresses the phase
noise by taking the difference between the phase measured at two antennas on the
same network interface card (NIC), referred to as the CSI phase difference. Such
data are decomposed by using the canonical polyadic decomposition, where the
decomposed signals are grouped and matched to a person with the use of ACF,
dynamic time warping, and the stable roommate matching algorithm. Having used
only the phase of the CSI, TensorBeat however does not address the blind-spot issue.

Building upon TensorBeat, PhaseBeat also uses the CSI phase difference
\cite{ref:phasebeat}. PhaseBeat, after performing calibration of the CSI phase
difference data, performs a subcarrier selection routine using the mean absolute
deviation. The discrete wavelet transform is applied to the selected subcarrier
to extract a respiratory waveform. Similar to TensorBeat, the blind-spot issue
has not been mitigated within the PhaseBeat algorithm.

Building upon TensorBeat and Phasebeat, Zhuo \emph{et al.}
\cite{ref:pos_free_breath}, propose using the ratio of the CSI measured by two
antennas on the same NIC. This CSI-ratio is then projected onto a line that
gives the largest breathing-to-noise ratio (BNR) plus variance on the complex
plane. This process is repeated for all subcarriers, each of which has its own
line for the CSI-ratios to project onto. Peak detection is used to align the
peaks along the projections, and PCA is applied with the first principal
component retained. The VMD algorithm is applied to the retained principal
component, decomposing the principal component into a set of intrinsic mode
functions (IMFs) that represent the oscillatory behaviour of the signal. The
respiratory waveform is assumed to be included in this set of IMFs and the IMF
with the largest variation is selected as the respiratory waveform. The
CSI-ratio addresses the blind-spot issue, while the combined PCA and VMD
strategy mitigates noise caused by non-ideal patient positions. We refer to this
method as Pos-Free Breath.

BreathTrack uses CSI phase differences together with a specialised hardware
configuration to mitigate phase noise \cite{ref:breath_track}. It is one of
the earliest CSI-based respiration monitoring methods to employ beamforming
for reducing multipath interference. BreathTrack estimates the dominant
multipath component using sparse recovery, incorporating both time-of-flight
and angle-of-arrival information. The phase of the estimated complex
coefficient is then used to derive the respiratory waveform, and the
respiratory rate is obtained from the strongest peak in the FFT spectrum.

Zeng \emph{et al.} \cite{ref:exp_mult_ant} propose a beamforming-based
approach that enhances the respiratory signal while suppressing phase noise
through target nulling. The genetic algorithm is used to determine
beamforming weights that minimise signal energy within the normal breathing
frequency range. These weights are then used to derive signal-enhancement
weights. The CSI-ratio between the signal-enhanced and target-nulled CSI is
used to cancel common phase noise. Finally, the respiratory waveform is
recovered using filtering and PCA, and the respiratory rate is estimated
through autocorrelation and peak detection \cite{ref:farsense}. We refer to
this method as Beamform.

\section{System Model}
\label{sec:system_model}

This section describes the underlying assumptions and signal model used
for extracting respiratory information from CSI data measured within a Wi-Fi
system. The proposed WiRM algorithm is designed for use with orthogonal frequency-division multiplexing (OFDM), assuming that the subcarrier bandwidth is much less than the system coherence bandwidth and that the symbol time is much less than the
system coherence time. These assumptions are well justified for indoor
environments and the application of respiratory rate monitoring. Here, the
relatively slow movements in the environments result in a large coherence time.
Furthermore, the typical root-mean-square delay spread for an indoor
environment is between $10$ and $1,000$ nanoseconds, resulting in a
coherence bandwidth (in the megahertz range) much larger than
the subcarrier bandwidths (in the kilohertz range) \cite{ref:andrea_goldsmith}.
Under these assumptions, the channel can be considered time-invariant over the
duration of a transmission symbol. Additionally, since the subcarrier bandwidth is much
smaller than the coherence bandwidth, each subcarrier can be modelled as
experiencing frequency-flat fading. 

WiRM requires a communication system with $A \geq 1$ transmit antennas on one
NIC and $B \geq 2$ receive antennas on another NIC. The NIC provides an estimate
of the time-dependent transfer function of the channel between the transmit
and the receive antennas, i.e., the CSI. Assume that
the CSI estimate provided by the NIC measured between a transmit antenna $T_a$ (for
${a \in \{1,2,\dotsc, A\}}$) and a receive antenna $R_b$ (for ${b \in
\{1,2,\dotsc,B\}}$) is free of interference from the other transmitted signals,
given a sufficient CSI estimation algorithm. 

The CSI measured when the environment is static is defined as
${H_{\text{s},T_a, R_b}(f) \in \mathbb{C}}$, where the subscript $T_a,R_b$
indicates that the variable is measured from the transmit antenna $T_a$ to the receive
antenna $R_b$. Assuming only one multipath component interacts with a person
breathing, we define ${\alpha_{T_a, R_b}(f) \in \mathbb{R}^{+}}$ to be the
magnitude attenuation, $\lambda$ the wavelength, ${D_{T_a, R_b}}$ the distance the multipath travels when the breathing is paused, ${\Delta
d_{\text{b}, T_a, R_b}}$ the depth of breathing, and ${\theta_{T_a, R_b}}$
the angle between the multipath and the vector tangent to the chest's
motion during a breath. The CSI measured between the transmitter $T_a$ and the receiver
$R_b$ is then defined as
\begin{align}
    \label{eq:system_model}
    H_{T_a, R_b}(t, f) & = H_{\text{s}, T_a, R_b}(f) + \alpha_{T_a, R_b}(f)
    \exp\Big(\frac{-j2\pi}{\lambda} \nonumber\\ & \qquad \times  \big(D_{T_a,
    R_b}  + \Delta d_{\text{b}, T_a, R_b}\sin(\theta_{T_a, R_b})r(t)\big)\Big),
\end{align} 
for each of the $K$ subcarrier center frequencies, ${f \in \{f_1, f_2, \dotsc, f_K\}}$. The respiratory waveform
$r(t) \in \mathbb{R}$ is periodic with an instantaneous frequency represented by ${B_{\text{Hz}}(t) \in
\mathbb{R}^{+}}$ or ${B(t) = 60B_\text{Hz}(t)}$ (beats per minute). While the system model in \eqref{eq:system_model} is adapted from TR-BREATH \cite{ref:tr_breath}, we use a general periodic function $r(t)$ instead of assuming a sinusoidal respiratory
waveform as in \cite{ref:tr_breath}.

The system model in \eqref{eq:system_model} is extended to include the effects
of common noises in CSI measurements as follows:
\begin{equation}
    \label{eq:non_ideal_system_model}
    \tilde{H}_{T_a, R_b}(t, f) = \kappa(t, f) H_{T_a, R_b}(t, f) e^{-j\eta(t, f)} + \epsilon(t, f), 
\end{equation}
where $\kappa(t,f) \in \mathbb{R}^+$ is log-normally distributed multiplicative
noise, $\epsilon(t,f) \in \mathbb{C}$ is additive white Gaussian thermal noise,
and $\eta(t, f) \in [-\pi, \pi]$ is uniformly distributed phase noise
\cite{ref:andrea_goldsmith}. With an abuse of notation, the system model will be
referred to in discrete-time for the remainder of the paper as $\tilde{H}_{T_a,
R_b}(n, f)$ for $n \in \{0, 1, \dotsc\}$, where $t = nT_\text{s}$ and
$T_\text{s}$ is the sampling period.

The task of respiration monitoring using CSI can be summarised as jointly
estimating the respiratory waveform $r(n)$ and the instantaneous frequency of
$r(n)$ (corresponding to the respiratory rate $B_\text{Hz}(n)$) from the
noise-infected ${\{\tilde{H}_{T_a, R_b}(n, f): a \in \{1,\dotsc, A\}, b \in
\{1,\dotsc, B\}\}}$ as measured by the NIC.

\subsection{Performance Criteria}
\label{ssec:performance_evaluation}

The proposed algorithm will be compared against current state-of-the-art
algorithms by using the average RMSE of the respiratory rate estimation
$\tilde{B}_\text{Hz}(n)$ versus the ground truth $B_\text{Hz}(n)$. Additionally,
we use the percentage of estimations within ${\pm3}$ BPM of the ground truth as
a comparative metric to assess how consistently each method approximates the
true respiratory rate.

To compare the respiratory waveform estimation, the absolute value of the
correlation between the estimated waveform $\tilde{r}(n)$ and the ground truth
$r(n)$ is computed. The absolute value is computed to account for if the
algorithm has estimated the correct waveform, but it is negated as
$\tilde{r}(n) = -r(n)$. The absolute correlation for a window of $\tilde N$ time
steps is computed via
\begin{equation}
    \label{eq:abs_correlation}
    |\rho_{r\tilde{r}}(n)| = \left | \frac{\sum^{n+\tilde{N}-1}_{i=n}(r(i) -
    \overline{r})(\tilde{r}(i) -
    \overline{\tilde{r}})}{\sqrt{\sum^{n+\tilde{N}-1}_{i=n}(r(i) -
\overline{r})^2}\sqrt{\sum^{n+\tilde{N}-1}_{i=n}(\tilde{r}(i) - \overline{\tilde{r}})^2}}\right |,
\end{equation}
where $\overline{r}$ and $\overline{\tilde{r}}$ are the sample mean of $r(i)$
and $\tilde{r}(i)$ for $i \in \{n, n+1, \dotsc, n+\tilde{N}-1\}$, respectively. The
correlation is computed over a sliding window, resulting in time-varying
correlation values $|\rho_{r\tilde{r}}(n)|$ for ${n\in\{0,1,\dotsc\}}$. The
average and maximum of the absolute correlation values are used to compare the
performance of each algorithm in estimating the respiratory waveform. 

\section{Proposed Wireless Respiration Monitoring (WiRM) Algorithm}
\label{sec:proposed_algorithm}

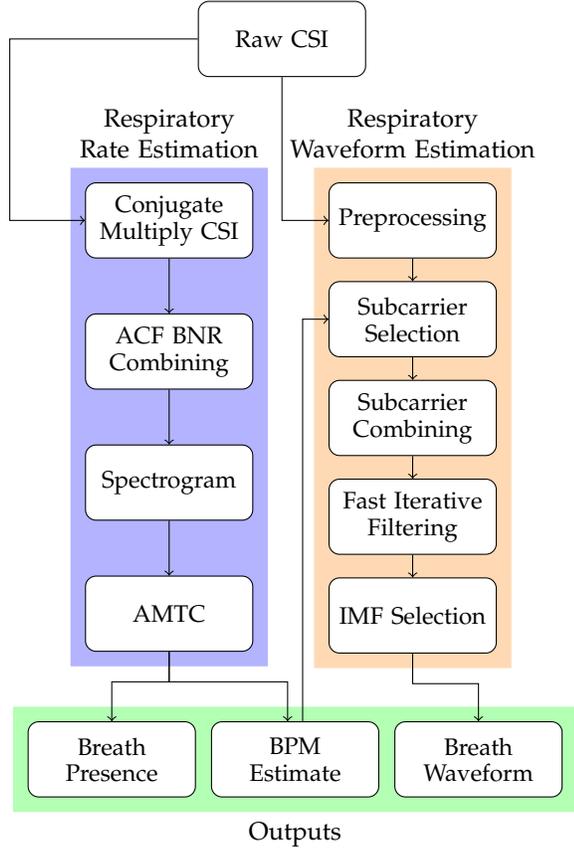
\begin{figure}
    \centering
    \scalebox{0.7}{
    \begin{tikzpicture}[node distance=0.4cm and 1cm]
        \node[flowchartnode] (raw_csi) {Raw CSI};

        \node (cm_compute) [flowchartnode, below=of raw_csi, yshift=-1cm,
        xshift=-1.5cm] {Conjugate Multiply CSI};
        \begin{scope}[node distance=0.73cm and 1cm]
        \node (acf_bnr) [flowchartnode, below=of cm_compute] {ACF BNR
        Combining};
        \node (spectrogram) [flowchartnode, below=of acf_bnr] {Spectrogram};
        \node (amtc) [flowchartnode, below=of spectrogram] {AMTC};
        \begin{pgfonlayer}{background}
            \node[draw=white, fill=blue!30, fit=(cm_compute) (acf_bnr) (spectrogram) (amtc), inner
            sep=0.2cm, label={[text width=4cm, align=center]above:Respiratory Rate Estimation}] {};
        \end{pgfonlayer}
        \end{scope}

        \begin{scope}[node distance=0.3cm and 1cm]
        \node (pre_process) [flowchartnode, right= of cm_compute]{Preprocessing};
        \node (subcarrier_selection) [flowchartnode, below=of pre_process] {Subcarrier Selection};
        \node (waveform_combine) [flowchartnode, below=of subcarrier_selection]
        {Subcarrier Combining};
        \node (fif) [flowchartnode, below=of waveform_combine] {Fast Iterative
        Filtering};
        \node (imf_select) [flowchartnode, below=of fif] {IMF
        Selection};
        \begin{pgfonlayer}{background}
            \node[draw=white, fill=orange!30,fit=(pre_process) (subcarrier_selection) (waveform_combine) (fif)
            (imf_select), inner
            sep=0.2cm, label={[text width=4cm, align=center]above:Respiratory
            Waveform Estimation}] {};
        \end{pgfonlayer}
        \end{scope}

        \node (detects_breathing) [flowchartnode, below=of imf_select, xshift=-4cm, yshift=-0.5cm] {Breath Presence};
        \node (bpm_estimate) [flowchartnode, right=of detects_breathing, xshift=-0.8cm] {BPM Estimate};
        \node (waveform) [flowchartnode, right=of bpm_estimate, xshift=-0.8cm] {Breath Waveform};

        \begin{pgfonlayer}{background}
            \node(outputs) [draw=white, fill=green!30, fit=(detects_breathing) (bpm_estimate) (waveform) , inner
            sep=0.2cm, label={[text width=4cm, align=center]below: Outputs}] {};
        \end{pgfonlayer}

        \node [coordinate, left=of cm_compute] (c1)  {};
        \draw [->] (raw_csi.west) -| (c1) -- (cm_compute.west);

        \draw [->] (cm_compute) -- (acf_bnr);
        \draw [->] (acf_bnr) -- (spectrogram);
        \draw [->] (spectrogram) -- (amtc);

        \node [coordinate, below=of amtc] (c2)  {};
        \draw [->] (amtc.south) |- (c2) -| (detects_breathing.north);

        \node [coordinate, left=0.1cm] (bpm_estimate_in) at (bpm_estimate.north) {};
        \draw [->] (amtc.south) |- (c2) -| (bpm_estimate_in);

        \node [coordinate, right=0.1cm] (bpm_estimate_out) at (bpm_estimate.north) {};
        \draw [->] (bpm_estimate_out) |- (subcarrier_selection);

        \draw [->] (raw_csi) |- (pre_process);
        \draw [->] (pre_process) -- (subcarrier_selection);
        \draw [->] (subcarrier_selection) -- (waveform_combine);
        \draw [->] (waveform_combine) -- (fif);
        \draw [->] (fif) -- (imf_select);
        \node [coordinate, below=of imf_select] (c3)  {};
        \draw [->] (imf_select) |- (c3) -| (waveform.north);
    \end{tikzpicture}
    }
    \vspace{\belowfig}
    \caption{Overview of the proposed WiRM algorithm.}
    \vspace{\belowfigcaption}
    \label{fig:system_overview}
\end{figure}
An overview of the proposed WiRM algorithm can be found in
Figure~\ref{fig:system_overview} the respiratory rate estimation block produces
an estimate of the current respiratory rate denoted $\tilde{B}_{\text{Hz}}(n)$,
and a boolean indicator $\tilde{\Phi}(n)$ that shows whether or not a breath was
detected within the CSI data for the current time step. The respiratory-rate
estimation procedure is described in
Subsection~\ref{ssec:respiratory_rate_estimation}. The output of the
respiratory-waveform estimation block in Figure~\ref{fig:system_overview} is an
estimate of the respiratory waveform denoted $\tilde{r}(n)$ as described in
Subsection~\ref{ssec:respiratory_waveform_estimation}.

One novelty of the proposed algorithm lies in the use of the respiratory rate
estimation to guide the respiratory waveform decomposition. This allows for an
easier search over the set of decompositions and provides more confidence that
the selected decomposed signal is indicative of breathing. While not a
definitive method for ensuring the decomposed signal is a breath, this
methodology greatly improves the decomposed signal's correlation with the ground
truth. The entire estimation procedure is outlined in
Figure~\ref{fig:system_overview}.

\subsection{Stage 1 -- Respiratory Rate Estimation}
\label{ssec:respiratory_rate_estimation}

\begin{figure}
    \centering
    \includesvg[width=0.7\columnwidth]{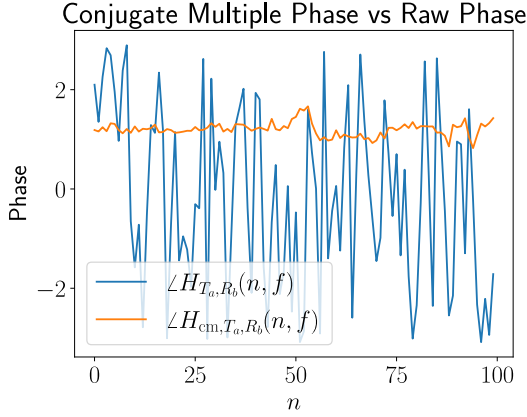}
    \vspace{\belowfig}
    \caption{
        The phase $\angle H_{T_a, R_b}(n,f)$ versus the conjugate multiplied phase
        $\angle H_{cm, T_a, R_b}(n,f)$. The figure shows improvements on the
        stability of the phase with the use of conjugate multiplication.
    }
    \vspace{\belowfigcaption}
    \label{fig:cm_phase_difference}
\end{figure}

The proposed respiratory rate estimation procedure can be found in
Figure~\ref{fig:system_overview}. Each block is summarised as follows.
\begin{enumerate}
    \item{
        The CSI data is preprocessed through conjugate multiplication. This
        nullifies phase noise within the CSI as shown in
        Figure~\ref{fig:cm_phase_difference} and detailed in
        Subsection~\ref{sssec:rr_pre_processing}.
    }
    \item{
        The ACF of the magnitude and phase of the conjugate multiplied CSI
        is computed. These ACFs contain peaks that arise due to periodic
        signals embedded within the conjugate multiplied CSI. The ACFs are
        combined to accentuate the peaks corresponding to breath signals and
        minimise noise within the signal. More details are available in Subsection~\ref{sssec:acf_bnr_combining}.
    }
    \item{
        To observe which frequencies are present within the ACF, the Fourier
        transform is applied to obtain the power spectral density function.
        This process is repeated over a sliding window, where each
        power spectral density function forms a column of a spectrogram.
        Creating this spectrogram over sliding windows allows for the
        frequencies within the ACF to be observed as they change over time.
        More details are available in Subsection~\ref{sssec:spectrogram}. 
    }
    \item{
        The adaptive multi-trace carving (AMTC) algorithm is finally applied to the spectrogram to achieve the
        final respiratory rate estimate. The AMTC algorithm uses transition
        probability functions from \eqref{eq:transition_prob} which are proposed
        from observations on empirical respiratory data. These transition
        probability functions incorporate the dynamics of breathing into the
        AMTC algorithm while tracking the respiratory rate. More details are available in Subsection~\ref{sssec:amtc}.
    }
\end{enumerate}

\subsubsection{Conjugate Multiple CSI}
\label{sssec:rr_pre_processing}

Phase noise such as sampling clock offset, carrier frequency offset, jitter, and
sampling time offset greatly impacts the reliability of CSI phase measurements.
Fortunately, these sources of noise are all common to antennas attached to the
same NIC and, equivalently, the same local oscillator. As such, we choose to
cancel out the phase noise $\eta(t, f)$ from \eqref{eq:non_ideal_system_model},
by conjugate multiplication \cite{ref:fullbreathe}. The conjugate multiplied CSI
data exist on the arc of an ellipse in the complex plane, moving along the arc
as the dynamic multipath changes \cite{ref:stat_sig_processing,
ref:fullbreathe}. That is, the conjugate multiplied CSI moves along the arc as
$r(t)$ changes in \eqref{eq:system_model}. Conjugate multiplication is chosen
for phase noise cancellation over the CSI-ratio method described in
Subsection~\ref{ssec:resp_waveform_lit_review}, as it preserves CSI magnitude
information that could otherwise be lost during the ratio operation.

CSI data are collected in windows of length $N$, where $N$ is chosen to ensure
the respiratory waveform remains approximately stationary while still
capturing at least one full breathing cycle. For each transmit antenna $T_a$ for
$a \in \{1, 2, \dotsc, A\}$ and receive antenna $R_b$ for ${b \in \{1, 2, 
\dotsc, B\}}$, $NK$ CSI values are collected, resulting in a total of $NKAB$
values. The conjugate multiplied CSI is computed by multiplying each CSI link
with the complex conjugate of a reference link from the same transmitter to a
fixed receiver $R_1$, as defined in
\begin{equation}
    \label{eq:cm_csi_calc}
    H_{\text{cm},T_a, R_b}(n,f) = H_{T_a,R_b}(n,f)\overline{H_{T_a,R_1}(n,f)}.
\end{equation}
This operation is performed for $a \in \{1, 2, \dotsc, A\}$ and ${b \in \{2, 3,
\dotsc, B\}}$, excluding the reference receiver $R_1$, resulting in $L_\text{cm}
= B(A-1)$ conjugate multiplied links. For brevity, we denote each of these
links as $H_{\text{cm}, i}$ where $i \in \{1,2,\dotsc, L_\text{cm}\}$.

For every $i$-th conjugate multiplied link, the magnitude and the phase of the
conjugate multiple CSI data are computed and represented by
$|H_{\text{cm},i}(n,f)|$ and $\angle H_{\text{cm} ,i}(n,f)$. 

\subsubsection{ACF BNR Combining}
\label{sssec:acf_bnr_combining}

The linear ACF is computed across time for each subcarrier and every conjugate
multiple link for both the magnitude $|H_{\text{cm}, i}(n, f)|$ and phase $\angle
H_{\text{cm}, i}(n,f)$. These ACFs are defined as 
\begin{equation}
    \label{eq:acf_of_carrier}
    \begin{split}
        \mathbf{R}'_{|\cdot|, \text{cm},i}(f) &= \begin{bmatrix}
            R_{|\cdot|, \text{cm},i}(0, f)\\
            R_{|\cdot|, \text{cm},i}(1, f)\\
            \vdots \\
            R_{|\cdot|, \text{cm},i}(N - 1, f)\\
        \end{bmatrix}\\
        \mathbf{R}'_{\angle, \text{cm}, i}(f) &= \begin{bmatrix}
            R_{\angle,\text{cm}, i}(0, f)\\
            R_{\angle,\text{cm}, i}(1, f)\\
            \vdots \\
            R_{\angle,\text{cm}, i}(N-1, f)\\
        \end{bmatrix},
    \end{split}
\end{equation}
where $\mathbf{R}'_{|\cdot|, \text{cm},i}(f)$ and $\mathbf{R}'_{\angle,
\text{cm}, i}(f)$ are unit-less column vectors denoting the ACF of the magnitude
and the ACF of the phase respectively for each subcarrier ${f \in
\{f_1,f_2,\dotsc, f_K\}}$ and conjugate multiple link ${i \in \{1,2,\dotsc,
L_{\text{cm}}\}}$. 

An important property of the ACF is that all white noise is correlated at a lag
of zero; thus, the $\tau = 0$ sample is omitted from the ACFs
$\mathbf{R}'_{|\cdot|, \text{cm},i}(f)$ and $\mathbf{R}'_{\angle, \text{cm},
i}(f)$ \cite{ref:smars}. This omission allows us to define

\begin{equation}
    \label{eq:acf_of_carrier_omit_tau_zero}
    \begin{split}
        \mathbf{R}_{|\cdot|, \text{cm},i}(f) &= \begin{bmatrix}
            R_{|\cdot|, \text{cm},i}(1, f)\\
            R_{|\cdot|, \text{cm},i}(2, f)\\
            \vdots \\
            R_{|\cdot|, \text{cm},i}(N - 1, f)\\
        \end{bmatrix}\\
        \mathbf{R}_{\angle, \text{cm}, i}(f) &= \begin{bmatrix}
            R_{\angle,\text{cm}, i}(1, f)\\
            R_{\angle,\text{cm}, i}(2, f)\\
            \vdots \\
            R_{\angle,\text{cm}, i}(N-1, f)\\
        \end{bmatrix},
    \end{split}
\end{equation}
for each subcarrier ${f \in \{f_1,f_2,\dotsc, f_K\}}$ and conjugate multiple link ${i
\in \{1,2,\dotsc, L_{\text{cm}}\}}$. Two matrices for each conjugate multiple
link are defined as 
\begin{equation}
    \label{eq:acf_power_phase}
    \begin{split}
        \mathbf{R}_{|\cdot|, \text{cm}, i} &= \Big[
            \mathbf{R}_{|\cdot|, \text{cm}, i}(f_1),\ 
            \mathbf{R}_{|\cdot|, \text{cm}, i}(f_2),\ 
            \dotsm,\ 
            \mathbf{R}_{|\cdot|, \text{cm}, i}(f_K) 
        \Big]\\
        \mathbf{R}_{\angle,\text{cm}, i} &= \Big[
            \mathbf{R}_{\angle,\text{cm}, i}(f_1),\ 
            \mathbf{R}_{\angle,\text{cm}, i}(f_2),\ 
            \dotsm,\ 
            \mathbf{R}_{\angle,\text{cm}, i}(f_K)
        \Big],
    \end{split}
\end{equation}
where both $\mathbf{R}_{|\cdot|, \text{cm}, i}$ and $\mathbf{R}_{\angle,
\text{cm}, i}$ are matrices with a dimension of $(N-1)\times K$ for each
conjugate multiple link ${i \in \{1, 2, \dotsc, L_{\text{cm}}\}}$.

With $\mathbf{R}_{|\cdot|, \text{cm}, i}$ and $\mathbf{R}_{\angle,\text{cm},i}$,
an augmented ACF matrix $\mathbf{R}_{H_{\text{cm}, i}}$ of dimension $(N - 1)\times 2K$ is defined as 
\begin{equation}
    \label{eq:aug_acf_one_link}
    \mathbf{R}_{\text{cm},i} = 
        \begin{bmatrix} 
            \mathbf{R}_{|\cdot|, \text{cm}, i},\ \mathbf{R}_{\angle, \text{cm}, i} 
        \end{bmatrix},
\end{equation}
for $i \in \{1, 2, \dotsc, L_{\text{cm}}\}$.

All of the ACF matrices for each conjugate multiple link are then further
augmented into one large matrix $\mathbf{R}_{H_{\text{cm}}}$ of dimension $(N-1) \times 2KL_{\text{cm}}$ defined as 
\begin{equation}
    \label{eq:aug_acf}
    \mathbf{R}_{\text{cm}} = \Big[
        \mathbf{R}_{\text{cm},1},\ 
        \mathbf{R}_{\text{cm},2},\ 
        \dotsm,\ 
        \mathbf{R}_{\text{cm},L_\text{cm}} 
    \Big].
\end{equation}

The phase-invariance property of the ACF means that any periodic signals
embedded within $|H_{\text{cm},i}(n,f)|$ and $\angle H_{\text{cm},i}(n,f)$ have
periods that correspond to the peaks found within the ACF. Importantly, if there
are common periodic signals in either $\mathbf{R}_{|\cdot|, \text{cm},i}(f)$ or
$\mathbf{R}_{\angle,\text{cm},i}(f)$ for all subcarriers ${f \in \{f_1, f_2,
\dotsc, f_K\}}$ and all conjugate multiple links ${i \in \{1,2,\dotsc,
L_{\text{cm}}\}}$, then the peaks corresponding to the common signal will align
along the columns. With this property, the high dimensionality of
$\mathbf{R}_{\text{cm}}$ and noise within the ACFs, a strategy is needed to combine all
of the ACFs (the columns of $\mathbf{R}_{\text{cm}}$). Such a strategy should
accentuate the peaks in the ACFs caused by breath and minimise noise.

Breathing-to-noise-ratio (BNR) combining is used to combine signals that
strengthens breath components while minimising frequency components outside of
the range of normal breathing \cite{ref:farsense}. The BNR
$\mathbb{B}(\mathbf{x})$ of a vector $\mathbf{x}$ is described as 

\begin{equation}
    \label{eq:bnr_definition}
    \mathbb{B}(\mathbf{x}) = \frac{\sum_{f \in \mathcal{B}}
    |\mathcal{F}(\mathbf{x})(f)|^2}{\sum_{f \in
    \mathcal{A}}|\mathcal{F}(\mathbf{x})(f)|^2},
\end{equation}
where $\mathcal{F}(\mathbf{x})$ is the FFT of $\mathbf{x}$, $\mathcal{A}$ the set
of all frequencies within the FFT of $\mathbf{x}$, and $\mathcal{B}$ a subset of
$\mathcal{A}$ corresponding to the normal respiratory rate. That is, ${\mathcal{B}
= \{f \in \mathcal{A} | f_\text{b,min} \leq f \leq f_\text{b,max}\}}$ where
$f_\text{b,min}$ and $f_\text{b,max}$ correspond to the minimum and maximum
normal respiratory rates, respectively \cite{ref:deep_breath}.

\begin{figure}
    \centering
    \includesvg[width=0.8\columnwidth]{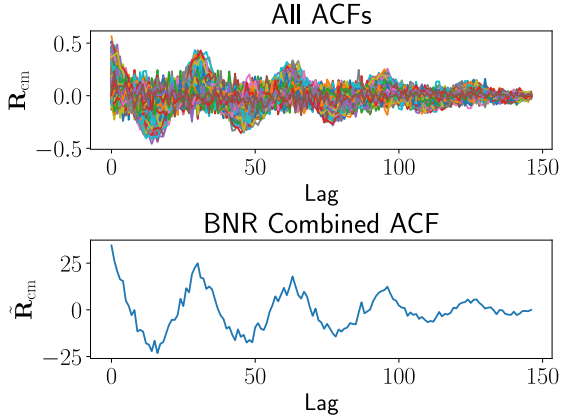}
    \vspace{\belowfig}
    \caption{
        All ACFs from $\mathbf{R}_{\text{cm}}$ versus the BNR combined ACF
        $\tilde{\mathbf{R}}_{\text{cm}}$ with a breath signal present.
    }
    \vspace{\belowfigcaption}
    \label{fig:bnr_acf}
\end{figure}

Each column of $\mathbf{R}_\text{cm}$ is weighted by its BNR, and the weighted
columns are then summed to give the combined ACF column vector
$\tilde{\mathbf{R}}_{\text{cm}}$ of length (N-1), defined as 
\begin{equation}
    \label{eq:bnr_combining}
    \tilde{\mathbf{R}}_{\text{cm}} = \sum_{j=0}^{2KL_{\text{cm}}}
    \mathbb{B}(\text{col}_j(\mathbf{R}_{\text{cm}}))\text{col}_j(\mathbf{R}_{\text{cm}}),
\end{equation}
where $\text{col}_j(\mathbf{R}_{\text{cm}})$ is the $j$-th column of
$\mathbf{R}_{\text{cm}}$ for ${j \in \{1, 2, \dotsc, 2KL_{\text{cm}}\}}$. As evident in Figure~\ref{fig:bnr_acf}, the
BNR combining significantly accentuates the breathing signal embedded within
$\mathbf{R}_\text{cm}$.

\subsubsection{Spectrogram}
\label{sssec:spectrogram}

By using the zoom FFT, the combined ACF $\tilde{\mathbf{R}}_{\text{cm}}$ is transformed into a power
spectral density to produce a smoother frequency representation that only focuses on the frequency range of human
breathing\cite{ref:zoom_fft_origin}. Note that the zoom FFT does not improve the frequency
resolution, but rather creates a smoother frequency representation, similar to zero-padding. 

A major disadvantage of using the power spectral density approach is due to
the Heisenberg uncertainty principle \cite{ref:heisenbeg_uncertainty_origin}. The resolution in the frequency domain is limited based on the length of the signal recording, which with the stationarity constraint on $N$ leads to poor
frequency resolution. All frequencies within a given frequency difference
$\Delta f = f_s / N$ (where $f_s$ is the sampling frequency) are merged into one
peak in the power spectral density plot, causing the poor
frequency resolution. All frequencies within this frequency region or bin are
sinc-interpolated, which, in practice, makes the maximum value of this peak
represent an average of all of the frequencies within the bin. This fact
allows the maximum value to still be useful as an approximate BPM estimate,
which, when combined with the technique described in Subsection~\ref{sssec:subcarrier-selection}, resolves the frequency resolution
issue.

To estimate both the breathing frequencies at the given time step $n$ for
the given window size $N$ and how the frequencies change over time, a
sliding window is employed with a total of $W$ sliding windows. Here, two
adjacent windows overlap by $Y$ samples. Denote the start and end times of the
$w$-th time window for ${w \in \{0,1, \dotsc, W-1\}}$ as ${w(N-Y)}$ and ${w(N-Y) + N - 1}$, respectively. The BNR-combined ACF (a length-$(N-1)$
column vector calculated according to \eqref{eq:bnr_combining}) for the $w$-th time window
is then denoted as $\tilde{\mathbf{R}}_{\text{cm}, w}$.

\begin{figure}
    \centering
    \includegraphics[width=0.85\columnwidth]{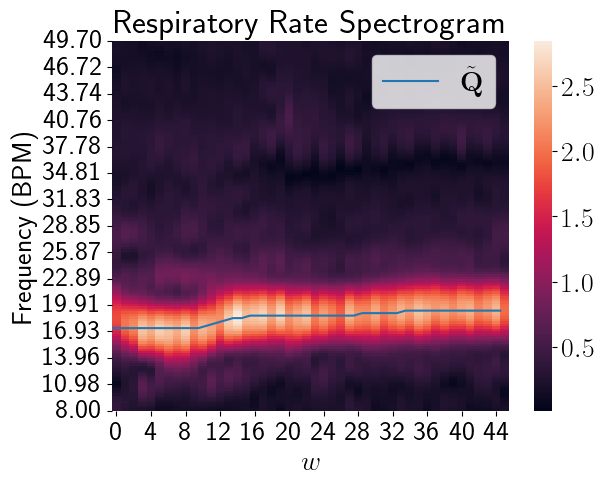}
    \vspace{\belowfig}
    \caption{
        The frequency spectrogram $\mathbf{S}$ displays a respiratory trace
        $\tilde{\mathbf{Q}}$ from the AMTC algorithm over time. Note that the
        frequency resolution issue is apparent in this image, with the width in
        frequency of the respiratory trace being visibly large.
    }
    \vspace{\belowfigcaption}
    \label{fig:frequency_spectrogram}
\end{figure}

A spectrogram $\mathbf{S}$ of dimension ${(N-1)\times W}$ is created by using
the sliding-window BNR-combined ACF $\tilde{\mathbf{R}}_{\text{cm}, w}$ for ${w
\in \{0,1,\dotsc, W-1\}}$ as follows:
\begin{equation}
    \label{eq:spectrogram_creation}
    \mathbf{S} = 
    \Big[
        \mathcal{F}_\text{z}(\tilde{\mathbf{R}}_{\text{cm}, 0}),\ 
        \mathcal{F}_\text{z}(\tilde{\mathbf{R}}_{\text{cm}, 1}),\ 
        \dotsm,\ 
        \mathcal{F}_\text{z}(\tilde{\mathbf{R}}_{{\text{cm}}, W-1})
    \Big],
\end{equation}
where the magnitude of the zoom FFT of a vector $\mathbf{x}$ is denoted as
$\mathcal{F}_\text{z}(\mathbf{x})$. The output of the zoom FFT is a column
vector of length $N - 1$ representing $N - 1$ evenly spaced frequencies from
$f_\text{b,min}$ to $f_\text{b,max}$. This results in the spectrogram
$\mathbf{S}$ being a matrix of dimension ${(N-1)\times W}$ with $N-1$
frequencies and $W$ time steps. Figure~\ref{fig:frequency_spectrogram} shows an example of a real breath signal as it evolves
over time within the spectrogram $\mathbf{S}$.

This procedure is very similar to the power spectral density generation
facilitated by the short-time Fourier transform \cite{ref:stft_origin}. However,
a major difference between the short-time Fourier transform and the approach
outlined in \eqref{eq:spectrogram_creation} is that the power spectral density
for each time window is computed from a combination of subcarriers, which could
change from one window to the next as described in \eqref{eq:bnr_combining}.

\subsubsection{AMTC}
\label{sssec:amtc}

The AMTC algorithm estimates a trace $\mathbf{Q}$ that describes how a frequency
changes over time within a spectrogram \cite{ref:amtc}. Here, we choose this
algorithm for its efficient computation and its ability to embed transition
probabilities into the estimation of frequency traces. The AMTC algorithm
assumes that frequency transitions from $Q(n-1)$ to the next frequency $Q(n)$
follows a Markov process. The exact probability distribution that describes the
likelihood of the frequency transition is system-dependent as:
\begin{equation}
    \label{eq:transition_prob}
    \begin{split}
        \log(P_\mathbf{Q}(\mathbf{q})) &= \log(P_{Q(0)}(q_0)) \\
                                       &\quad+ \sum^{W-1}_{w=1}
                                       \log(P_{Q(w)|Q(w-1)}(q_w|q_{w-1})),
    \end{split}
\end{equation}
where $\mathbf{q} = [q_0, q_1, \dotsc, q_{W-1}]$, $P_{Q(0)}(q_0)$ is the
probability of the first frequency $Q(0)$ occurring, and
$P_{Q(w)|Q(w-1)}(q_w|q_{w-1})$ the conditional probability of the
frequency $Q(w)$ occurring given that the previous frequency is ${Q(w-1)}$ for
${Q(w) \in \{f_1, f_2, \dotsc, f_K\}}$ and ${w \in \{0,1, \dotsc, W-1\}}$.

Using the AMTC algorithm, we aim to jointly estimate the respiratory rates ${Q(w) \in
\{f_1, f_2, \dotsc, f_{N-1}\}}$ for all sliding time windows ${w \in \{0,1,\dotsc,
W-1\}}$ from the spectrogram $\mathbf{S}$. Define the estimated respiratory rate trace as ${\tilde{\mathbf{Q}}=
[\tilde{Q}(0), \tilde{Q}(1), \dotsc, \tilde{Q}(W-1)]}$. For the purposes of respiratory rate tracking, we assume $P_{Q(1)}(q_1)$ and
$P_{Q(w)|Q(w-1)}(q_w|q_{w-1})$ from \eqref{eq:transition_prob} as a uniform distribution and
a zero-mean Gaussian distribution, respectively as: 
\begin{equation}
    \label{eq:breath_distribution}
    \begin{split}
        P_{Q(0)}(q_1) &= \frac{1}{N - 1}, \\
        P_{Q(w)|Q(w-1)}(q_w|q_{w-1}) &= \frac{1}{\sigma
        \sqrt{2\pi}}e^{-\frac{(q_w - q_{w-1})^2}{2\sigma^2}},
    \end{split}
\end{equation}
for all $w \in \{0,1,\dotsc, W-1\}$ where $\sigma$ is the standard deviation of
the Gaussian distribution.

The choice of a uniform distribution is justified by the lack of prior knowledge
of what the respiratory rate may be at $w = 0$ and, as such, all frequencies
${Q(0) \in \{f_1, f_2, \dotsc, f_{N-1}\}}$ are assumed equally likely. The
zero-mean Gaussian distribution is proposed based on evidence found during the
analysis of a respiratory waveform dataset \cite{ref:respiratory_waveforms}.
Here, the respiratory waveform data is broken into windows in the same manner as
required by WiRM. The respiratory rates were computed for each time window, with
the difference between subsequent rates being calculated as the frequency
transition. Assembling these breathing frequency transitions into a histogram
strongly resembled a zero-mean Gaussian distribution. Curve fitting suggested an
empirical value of $\sigma = 4.552$ for the standard deviation of this Gaussian
distribution.

The AMTC algorithm produces a frequency trace of length $W$ by finding the trace
that maximises the energy along the trace from $\mathbf{S}$, rewarded by the
frequency transition probability. Maximising the summation of the energy and the
frequency transition probability allows for domain-specific applications of the
AMTC algorithm, where the frequency transition probability can be chosen to
better suit specific scenarios, as we have demonstrated in
\eqref{eq:breath_distribution}. The estimated frequency trace is calculated as 
\begin{equation}
    \label{eq:amtc_freq_trace}
    \tilde{\mathbf{Q}} \in \operatorname*{argmax}_{\tilde{\mathbf{q}}}
    \Big(E(\tilde{\mathbf{q}}) + \Lambda \log(P_\mathbf{Q}(\tilde{\mathbf{q}}))\Big),
\end{equation}
where $E(\tilde{\mathbf{q}})$ is the total energy of the trace taken from
$\mathbf{S}$, $\Lambda \in \mathbb{R}^+$ is a tuneable smoothing parameter, and
${\tilde{\mathbf{q}} \in \{f_1, f_2, \dotsc, f_{N-1}\}^{N-1}}$. Ties in $\tilde{\mathbf{Q}}$ are broken arbitrarily. 

The respiratory rate estimation for the current timestep $\tilde{B}_{\text{Hz}}(n)$
is finally computed as
\begin{equation}
    \label{eq:rate_computation}
    \tilde{B}_{\text{Hz}}(n) =
    \tilde{Q}\left(\doubleceil{\frac{n - N + 1}{N - Y}}\right),
\end{equation}
for ${n \in \{0, 1, \dotsc, W(N-Y) + N - 1\}}$ and where $\doubleceil{x} = \max(0, \lceil x\rceil)$.

Along with the frequency trace $\tilde{\mathbf{Q}}$, the AMTC algorithm also
returns the trace ${\mathbf{\Phi} = [\Phi(0), \Phi(1), \dotsc, \Phi(W-1)]}$,
where $\Phi(w)$ equals $1$ if breath is detected in the $w$-th window and $0$
otherwise \cite{ref:amtc}. Breathing detection for each time step is computed
similarly to \eqref{eq:rate_computation}. 

Algorithm~\ref{algo:resp_rate_estimation} summarises the proposed respiratory rate estimation algorithm, where $\text{AMTC}(\mathbf{S})$ denotes performing the AMTC algorithm on the
spectrogram $\mathbf{S}$, and $\text{col}_w(\mathbf{S})$ denotes the $w$-th
column of $\mathbf{S}$. 
\begin{algorithm}
    \caption{WiRM Stage 1 -- Respiratory Rate Estimation}
    \label{algo:resp_rate_estimation}

    \Input{$W$ time windows of $N$ samples of CSI data $H_{T_a, R_b}(n, f)$ for
    $T_a, \ a \in \{1, 2, \dotsc, A\}$ and $R_b, \ b \in \{1, 2, \dotsc, B\}$}
    \Output{The current BPM estimate $\tilde{B}_\text{Hz}(n)$ and the boolean
    indicator of breath presence $\tilde{\Phi}(n)$}

    $\mathbf{S} \gets \text{empty 2D matrix of dimension }(N-1)\times W$;

    \For {$w \in \{1,2, \dotsc, W\}$}{
        compute $H_{\text{cm}, i}(n,f)$ for $i \in \{1,2,\dotsc, L_\text{cm}\}$ via \eqref{eq:cm_csi_calc};

        compute $\mathbf{R}_{|\cdot|, \text{cm}, i}(f)$ and
        $\mathbf{R}_{\angle, \text{cm}, i}(f)$ via \eqref{eq:acf_of_carrier_omit_tau_zero};

        compute $\mathbf{R}_{\text{cm}}$ via \eqref{eq:aug_acf_one_link} and
        \eqref{eq:aug_acf};

        compute $\tilde{\mathbf{R}}_{\text{cm}}$ via \eqref{eq:bnr_combining};

        $\text{col}_w(\mathbf{S}) \gets \tilde{\mathbf{R}}_{\text{cm}}$;
    }

    $\tilde{\mathbf{Q}}, \mathbf{\Phi} \gets \text{AMTC}(\mathbf{S})$;

    $\tilde{B}_{\text{Hz}}(n) \gets \tilde{Q}(\left(\doubleceil{\frac{n - N + 1}{N - Y}}\right))$;

    $\tilde{\Phi}(n) \gets \Phi\left(\doubleceil{\frac{n - N + 1}{N - Y}}\right)$;

    \Return{$\tilde{B}_{\text{Hz}}(n), \tilde{\Phi}(n)$}

\end{algorithm}

\subsubsection{Computational Complexity}
\label{sssec:resp_rate_comp_complexity}

The computational complexity of the proposed respiratory rate estimation
algorithm is shown in Table~\ref{tab:resp_rate_comp_complexity}. The
proposed algorithm requires that enough CSI data has already been processed
to compute the spectrogram $\mathbf{S}$ before an estimation can be
produced. After this, the algorithm will process a new sliding window of
length $N$ for the next respiratory rate estimate as described in
Subsection~\ref{sssec:spectrogram}, the reported computational complexity in
Table~\ref{tab:resp_rate_comp_complexity} is this amortised result. 

\begin{table}[tb!]
    \vspace{\abovetable}
    \caption{Respiratory Rate Estimation Computational Complexity}
    \label{tab:resp_rate_comp_complexity}
    \vspace{\betweentablecaption}
    \centering
    \begin{tabular}{@{} l l @{}}
        \toprule
        Subsection & Complexity \\
        \midrule
        \ref{sssec:rr_pre_processing} & $O(NKL_\text{cm})$ \\ 
        \ref{sssec:acf_bnr_combining} & $O(NKL_\text{cm}\log(N))$\\
        \ref{sssec:spectrogram} & $O(N\log(N))$ \\
        \ref{sssec:amtc} & $O(WN^2)$ \cite{ref:amtc} \\
        \bottomrule
    \end{tabular}
    \vspace{\belowtable}
\end{table}

\subsection{Stage 2 -- Respiratory Waveform Estimation}
\label{ssec:respiratory_waveform_estimation}
The proposed respiratory waveform estimation procedure can be found in
Figure~\ref{fig:system_overview}. Each block is summarised as follows.
\begin{enumerate}
    \item{
        The raw CSI data is pre-processed similarly to
        Subsection~\ref{sssec:rr_pre_processing}; however, a much longer time
        period is measured, where the non-stationary dynamics of breathing can
        be observed. This is detailed in
        Subsection~\ref{sssec:waveform_pre_process}.
    }
    \item{
        The subcarrier that most strongly contains the respiratory rate
        estimation from Subsection~\ref{sssec:amtc} is selected as the primary
        subcarrier, as detailed further in Subsection~\ref{sssec:subcarrier-selection}.
    }
    \item{
        The primary subcarrier may still contain noise and so a combining
        procedure is proposed. More details are available in Subsection~\ref{sssec:subcarrier_combining}.
    }
    \item{
        The fast iterative filtering algorithm is employed to decompose the
        combined subcarrier into a set of IMFs, one of which is the estimated
        respiratory waveform. More details are available in Subsection~\ref{sssec:fif}.
    }
    \item{
        The respiratory waveform is selected from the set of IMFs as further detailed
        in Subsection~\ref{sssec:imf_selection}, giving the final estimated
        respiratory waveform $\tilde{r}(n)$.
    }
\end{enumerate}

\subsubsection{Pre-Processing}
\label{sssec:waveform_pre_process}

As previously noted in Subsection~\ref{sssec:spectrogram}, the frequency resolution of the
FFT is directly determined by the duration of the observed signal. Specifically,
shorter observation windows result in poorer frequency resolution, while longer
windows yield better frequency resolution. However, in the context of
respiration monitoring, longer observation periods increase the likelihood of
changes in the respiratory rate, resulting in non-stationary behaviour.
Fortunately, iterative filtering allows for the separation of such
non-stationary signals \cite{ref:fif}.

To improve the ability of iterative filtering to distinguish signals of similar
frequency from each other (i.e., improve the frequency resolution), a window of
$\tilde{N}$ CSI samples are collected where $\tilde{N} > N$
\cite{ref:fif_one_or_two_freq}. Recall that $N$ was chosen such that the
respiratory rate could be considered stationary, so this constraint does not exist
on $\tilde{N}$. The $\tilde{N}$ samples of raw CSI data is conjugate multiplied in \eqref{eq:cm_csi_calc}
to produce $H_{\text{cm}, i}(n,f)$ for conjugate multiple link ${i \in
\{1,2,\dotsc, L_{\text{cm}}\}}$, time ${n \in \{0, 1, \dotsc, \tilde{N} - 1\}}$,
and frequency ${f \in \{f_1, f_2, \dotsc, f_K\}}$.

The magnitude and phase of the conjugate multiplied CSI data are computed and
$z$-score normalised to make them unitless signals so that we can combine them for noise removal in
Subsection~\ref{sssec:subcarrier_combining}. The $z$-score normalisation was not
previously required in Subsection~\ref{ssec:respiratory_rate_estimation} since the ACF
already eliminates the units of the signal. The $z$-score
normalisation of a vector $\mathbf{x}$ is defined as 
\begin{equation}
    \label{eq:z_score_norm}
    z(\mathbf{x}) = \frac{\mathbf{x} - \mu_{\mathbf{x}}}{\sigma_{\mathbf{x}}},
\end{equation}
where $\mu_{\mathbf{x}}$ is the mean of $\mathbf{x}$, and $\sigma_\mathbf{x}$ is
the standard deviation. The subtraction $\mathbf{x} - \mu_{\mathbf{x}}$ denotes
the element-wise subtraction of the scalar mean from each element of the vector
$\mathbf{x}$.

The $\tilde{N}\times K$ dimension augmented matrices
$\mathbf{Z}_{|\cdot|, \text{cm}, i}$ and $\mathbf{Z}_{\angle, \text{cm}, i}$ are defined as 
\begin{align}
    \label{eq:aug_waveform_csi}
    \mathbf{Z}_{|\cdot|, \text{cm}, i} =
        \Big[&z\left(|H_{\text{cm}, i}(n,f_1)|\right), z\left(|H_{\text{cm},
            i}(n,f_2)|\right), \dotsm, \nonumber \\
             &z\left(|H_{\text{cm}, i}(n,f_K)|\right) \Big] \nonumber \\
    \mathbf{Z}_{\angle, \text{cm}, i} =
        \Big[&z\left(\angle H_{\text{cm}, i}(n,f_1)\right), z\left(\angle
            H_{\text{cm}, i}(n,f_2)\right), \dotsm, \nonumber \\
             &z\left(\angle H_{\text{cm}, i}(n,f_K)\right) \Big] 
\end{align}
for conjugate multiple link ${i \in \{1,2,\cdots, L_{\text{cm}}\}}$ and time ${n
\in \{0, 1, \cdots, \tilde{N} - 1\}}$. These matrices are then further augmented
into one large matrix as
\begin{align}
        \label{eq:z_score_augmented}
        \mathbf{Z}_{\text{cm}} = \Big[ &\mathbf{Z}_{|\cdot|, \text{cm}, 1},
        \mathbf{Z}_{\angle, \text{cm}, 1},
        \mathbf{Z}_{|\cdot|, \text{cm}, 2},
        \mathbf{Z}_{\angle, \text{cm}, 2},
        \dotsm,\nonumber\\ 
        &\mathbf{Z}_{|\cdot|, \text{cm}, L_\text{cm}},
    \mathbf{Z}_{\angle, \text{cm}, L_\text{cm}}\Big]
\end{align}
which forms the $\tilde{N} \times 2KL_\text{cm}$ matrix
$\mathbf{Z}_{\text{cm}}$.

\subsubsection{Subcarrier Selection}
\label{sssec:subcarrier-selection}

The subcarrier selection procedure is performed using the respiratory estimate
from Subsection~\ref{sssec:amtc}. As new length-$N$ time windows are processed,
a sliding AMTC algorithm is performed to compute a respiratory estimate for the last time
window. The respiratory estimate for this last time window is used in
the subcarrier selection procedure. 

We propose that the subcarrier magnitude or phase (i.e., the columns of
$\mathbf{Z}_{\text{cm}}$) that contains a strong frequency component at the respiratory rate $\tilde{B}_\text{Hz}$ within the last $N$
samples is chosen as the primary subcarrier. If we denote the element within
$\mathbf{Z}_{\text{cm}}$ at the $i$-th row and the $j$-th column as $\left
(\mathbf{Z}_{\text{cm}}\right)_{i, j}$, then the discrete-time Fourier transform
of each column of $\mathbf{Z}_\text{cm}$ evaluated at the breathing frequency
$\tilde{B}_\text{Hz}$ is computed as 
\begin{equation}
    \label{eq:dtft_subcarriers}
    \begin{bmatrix}
        F_{1}\\
        F_{2}\\
        \vdots\\
        F_{2KL_\text{cm}}
    \end{bmatrix}^\intercal
     = \begin{bmatrix}
        \sum^{\tilde{N}-1}_{n =\tilde{N} - 1 -N} \left
    (\mathbf{Z}_{\text{cm}}\right)_{n, 1} e^{-j2\pi \tilde{B}_\text{Hz}nT_s}\\
        \sum^{\tilde{N}-1}_{n =\tilde{N} - 1 -N} \left
    (\mathbf{Z}_{\text{cm}}\right)_{n, 2} e^{-j2\pi \tilde{B}_\text{Hz}nT_s}\\
        \vdots\\
        \sum^{\tilde{N}-1}_{n =\tilde{N} - 1 -N} \left
    (\mathbf{Z}_{\text{cm}}\right)_{n, 2KL_\text{cm}} e^{-j2\pi \tilde{B}_\text{Hz}nT_s}
\end{bmatrix}^\intercal,
\end{equation}
where ${[F_1, F_2, \dotsc, F_{2KL_\text{cm}}]}$ is a row vector denoting the
discrete-time Fourier transform evaluated at a frequency of
$\tilde{B}_\text{Hz}$ for the last $N$ elements of each column of
$\mathbf{Z}_\text{cm}$, and $T_s$ is the sample period.

The index of the primary subcarrier $c_\text{ps}$ is determined as
\begin{equation}
    \label{eq:primary_subcarrier_idx}
    c_\text{ps} \in \operatorname*{argmax}_c
    \left(|F_c| \right)
\end{equation}
for ${c \in \{1,2,\dotsc, 2KL_\text{cm}\}}$. Ties in $c_\text{ps}$ are broken
by selecting the first index that gives the maximum magnitude $|F_c|$, yielding
a single value $c_\text{ps}$. We can extract the
magnitude and phase of the primary subcarrier as $|F_{c_\text{ps}}|$ and $\angle F_{c_\text{ps}}$,
respectively. 

This entire process is equivalent to taking the magnitude of the discrete-time
Fourier transform of the last $N$ samples of each subcarrier's magnitude and
phase, evaluated at the frequency of breathing and choosing the largest
magnitude as the primary subcarrier. 

\subsubsection{Subcarrier Combining}
\label{sssec:subcarrier_combining}
Even though the primary subcarrier contains a strong frequency component at the respiratory rate $\tilde{B}_\text{Hz}$, there may still
be significant noise present. As such, we now present a combining strategy that is a variant
of the breathing-to-noise ratio combining strategy presented in
Subsection~\ref{sssec:acf_bnr_combining}. From \eqref{eq:system_model}, it can be
seen that the breathing signal must be in phase or $180^{\circ}$ out of phase
across subcarriers. An appropriate phase shift (either $0^\circ$ or $180^\circ$)
can be chosen and applied to each subcarrier to align it with the primary
subcarrier. The phase shift can be achieved by either negating the
subcarrier or not. Whether or not a subcarrier must be negated can be determined through \eqref{eq:dtft_subcarriers} and \eqref{eq:primary_subcarrier_idx} as
\begin{equation}
    \label{eq:combining_phase_shift}
    \beta_c = \begin{cases}
        1, & |\angle F_{c_\text{ps}} - \angle F_{c}| \leq \frac{\pi}{2}\\
        -1, & \text{otherwise}
    \end{cases},
\end{equation}
for $c \in \{1,2,\dotsc, 2KL_\text{cm}\}$. The insight here is that if a given subcarrier contains the frequency of breathing approximately in phase with the primary subcarrier, then the absolute difference between the phase at the breath frequency $\tilde{B}_\text{Hz}$ should at least be less than $\pi/2$.

Additionally, each subcarrier during the combining process can be scaled by their
relative magnitudes to the primary subcarrier. Doing so ensures that subcarriers
that contain a strong frequency component at $\tilde{B}_\text{Hz}$ are
given higher weights than subcarriers that do not. The overall combining
procedure combines the columns of $\mathbf{Z}_\text{cm}$ into a combined
subcarrier column vector $\tilde{\mathbf{p}}$ of length $\tilde{N}$
as
\begin{equation}
    \label{eq:subcarrier_combining}
    \tilde{\mathbf{p}} = \sum^{2KL_\text{cm}}_{c=1} \beta_c
    \frac{|F_c|}{|F_{c_\text{ps}}|}\text{col}_c\left(\mathbf{Z}_\text{cm}\right), 
\end{equation}
where $\text{col}_c\left(\mathbf{Z}_\text{cm}\right)$ denotes the
$c$-th column of $\mathbf{Z}_\text{cm}$.

\subsubsection{Fast Iterative Filtering}
\label{sssec:fif}
The combined subcarrier $\tilde{\mathbf{p}}$ contains a strong component at the respiratory rate frequency. Unfortunately, it may also still contain noise such as that from non-respiration motion or environment. To produce an estimate of the
respiratory waveform, the combined subcarrier can be decomposed into a set of
IMFs where, if the decomposition was successful, one of the IMFs is the
estimated respiratory waveform. Given the frequency resolution constraints and the non-stationary dynamics of
respiration as described in Subsection~\ref{sssec:waveform_pre_process}, the fast
iterative filtering (FIF) algorithm is chosen to perform the decomposition
\cite{ref:if_originator,ref:fif}. FIF algorithm has proven convergence, efficiency, and ability
to decompose similar non-stationary frequencies \cite{ref:fif_one_or_two_freq}.

The FIF algorithm decomposes a vector $\tilde{\mathbf{p}}$ into a matrix with dimension $\tilde{N}\times
D$, where $D$ is the number of IMFs generated by the FIF algorithm. $D$ depends
on the oscillatory modes of $\tilde{\mathbf{p}}$ and is varying. The
application of FIF to $\tilde{\mathbf{p}}$ is expressed as
\begin{equation}
    \label{eq:fif_decomposition}
    \text{FIF}(\tilde{\mathbf{p}}) = \Big[
        \text{IMF}_1,\ 
        \text{IMF}_2,\ 
        \dotsm,\ 
        \text{IMF}_D
    \Big],
\end{equation}
where $\text{IMF}_i$ is a length-$\tilde{N}$ column vector denoting the $i$-th
IMF for ${i \in \{0,1,\dotsc, \tilde{D}\}}$.

FIF can potentially suffer from mode-splitting, i.e., oscillatory modes from one physical phenomenon spread across multiple IMFs
\cite{ref:fif_best_practices}. It produces a
filter width of
\begin{equation}
    \label{eq:fif_filter_width}
    l = 2\left\lfloor\chi\frac{m}{k}\right\rfloor,
\end{equation}
where $m$ is the number of points in the signal, $k$ the number of extrema
points in the signal, and $\chi$ a tuning parameter. The value of $\chi$ must be
tuned to avoid mode-splitting and is chosen according to
Section~\ref{sec:results}. The tuning parameter $\chi$ is empirically determined
by selecting the value that maximises the average absolute correlation between
the decomposed respiratory waveform and the ground truth. The value of $\chi$
affects the oscillatory modes that are decomposed from a given signal. After
initially tuning the value of $\chi$, given similar waveform characteristics
such as breathing patterns, comparable frequency, and shape, this tuning
parameter should not require further refinement.

\subsubsection{IMF Selection}
\label{sssec:imf_selection}

The respiratory waveform must now be selected from the matrix of IMFs
$\text{FIF}(\tilde{\mathbf{p}})$. To this end, we employ a similar technique
to Subsection~\ref{sssec:subcarrier-selection}. Given that we have a respiratory
rate estimate $\tilde{B}_\text{Hz}$, which is the respiratory rate that is
computed from the previous $N$ time steps, we propose that the IMF with its
dominant frequency component closest to $\tilde{B}_\text{Hz}$ is selected as
the respiratory waveform.

If we define a column vector containing the last $N$ elements of $\text{IMF}_i$
as $\tilde{\text{IMF}}_i$ for $i \in \{1, 2, \dotsc, D\}$, then the dominant
frequency component is computed as
\begin{equation}
    \label{eq:peak_freq_ith}
    f_{\text{peak},i} \in \operatorname*{argmax}_f \left(\text{row}_f(|\mathcal{F}_\text{z}(\tilde{\text{IMF}}_i)|)\right)
\end{equation}
for $i \in \{1,2,\dotsc, D\}$. Here, $\text{row}_i(\mathbf{x})$ denotes the
$i$-th row of a column vector $\mathbf{x}$, and ${f_{\text{peak}, i} \in
[f_{\text{b,min}}, f_{\text{b,max}}]}$ is the dominant frequency component with
ties within $\operatorname*{argmax}_f$ being broken by choosing the first
value. 

The index $i_\text{b}$ of the IMF that has its peak closest to the respiratory
waveform frequency $\tilde{B}_\text{Hz}$ is computed as 
\begin{equation}
    \label{eq:imf_index}
    i_\text{b} \in \operatorname*{argmin}_i
    \left (\left |f_{\text{peak}, i} - \tilde{B}_\text{Hz}\right |\right )
\end{equation}
for $i \in \{1,2, \dotsc, D\}$. Here, ties in $i_\text{b}$ are broken arbitrarily yielding a single value $i_\text{b}$.

\begin{figure}
    \centering
    \includesvg[width=0.8\columnwidth]{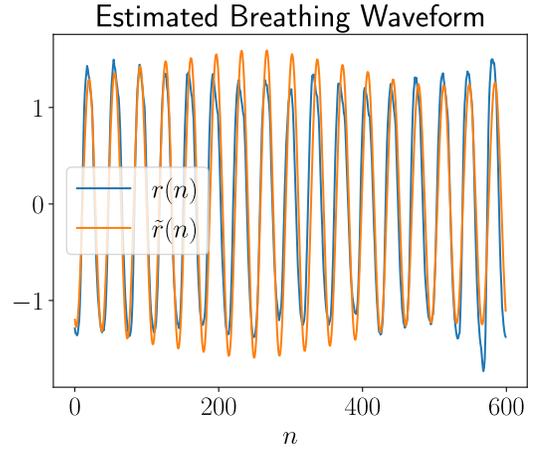}
    \vspace{\belowfig}
    \caption{
        The estimated breath waveform $\tilde{r}(n)$ compared to the actual
        respiratory waveform $r(n)$ measured during a polysomnography test.
    }
    \vspace{\belowfigcaption}
    \label{fig:estimated_waveform}
\end{figure}

Finally, the respiratory waveform is estimated as 
\begin{equation}
    \label{eq:final_imf_selection}
    \tilde{\mathbf{r}} = \text{col}_{i_\text{b}}(\text{FIF}(\tilde{\mathbf{p}})).
\end{equation}
This selected IMF is the final estimated respiratory waveform representing the
motion of the chest as a person breathes, as shown in
Figure~\ref{fig:estimated_waveform}. This final waveform of length $\tilde{N}$
is a non-stationary signal that appropriately captures the dynamics of a person's
breath allowing for its use in medical scenarios.

Algorithm~\ref{algo:waveform_estimation} summarises the respiratory waveform estimation algorithm.

\begin{algorithm}
    \caption{WiRM Stage 2 -- Respiratory Waveform Estimation}
    \label{algo:waveform_estimation}

    \Input{
        $L_\text{cm}K\tilde{N}$ CSI measurements $H(n, f)$, and the current respiratory rate
    estimate in Hertz $\tilde{B}_{\text{Hz}}(n)$.
    }
    \Output{The respiratory waveform estimate $\tilde{r}(n)$.}

    \For {link $i \in \{1,2,\dotsc, L_\text{cm}\}$}{

        compute $\mathbf{Z}_{|\cdot|, \text{cm}, i}$ and
        $\mathbf{Z}_{\angle,\text{cm},i}$ as in \eqref{eq:aug_waveform_csi};
    }

    form $\mathbf{Z}_{\text{cm}}$ as in \eqref{eq:z_score_augmented};

    compute $[F_1, F_2, \dotsc, F_{2KL_\text{cm}}]$ as in \eqref{eq:dtft_subcarriers};

    find the primary subcarrier index $c_\text{ps}$ as in \eqref{eq:primary_subcarrier_idx};

    compute the combined subcarrier $\tilde{\mathbf{p}}$ as in \eqref{eq:subcarrier_combining};

    compute $\text{FIF}(\tilde{\mathbf{p}})$ as in \eqref{eq:fif_decomposition};

    find the frequency peaks of each IMF, $f_{\text{peak},i}$ for $i \in \{1,2,\dotsc, D\}$, as in \eqref{eq:peak_freq_ith};

    find the IMF index $i_\text{b}$ with the closest frequency peak to $\tilde{B}_\text{Hz}(n)$
    as in \eqref{eq:imf_index};

    $\tilde{\mathbf{r}} \gets \text{col}_{i_\text{b}}(\text{FIF}(\tilde{\mathbf{p}}))$;

    \Return{$\tilde{\mathbf{r}}$}
\end{algorithm}

\subsubsection{Computational Complexity}
\label{sssec:resp_waveform_comp_complexity}

The computational complexity of the proposed respiratory waveform estimation
algorithm is shown in Table~\ref{tab:resp_waveform_comp_complexity}. As with the
computational complexity shown in
Subsection~\ref{sssec:resp_rate_comp_complexity}, we assume that sufficient data
has been pre-processed such that the computational complexity is the
amortised cost.

\begin{table}[tb!]
    \vspace{\abovetable}
    \caption{Respiratory Waveform Estimation Computational Complexity}
    \label{tab:resp_waveform_comp_complexity}
    \vspace{\betweentablecaption}
    \centering
    \begin{tabular}{@{} l l @{}}
        \toprule
        Subsection & Complexity \\
        \midrule
        \ref{sssec:waveform_pre_process} & $O(\tilde{N}KL_\text{cm})$\\ 
        \ref{sssec:subcarrier-selection} & $O(NKL_\text{cm})$\\
        \ref{sssec:subcarrier_combining} & $O(\tilde{N}KL_\text{cm})$\\
        \ref{sssec:fif} & $O(\tilde{N}\log(\tilde{N}))$ \cite{ref:mvfif}\\
        \ref{sssec:imf_selection} & $O(\tilde{N}(D + \log(\tilde{N})))$\\
        \bottomrule
    \end{tabular}
    \vspace{\belowtable}
\end{table}

\section{Performance Evaluation} 
\label{sec:results}
The performance of the proposed algorithm WiRM is evaluated and compared against
four state-of-the-art methods, namely, Pos-Free Breath
\cite{ref:pos_free_breath}, Beamform \cite{ref:exp_mult_ant}, TR-BREATH
\cite{ref:tr_breath}, and SMARS \cite{ref:smars}. The three criteria for
comparison are (i) the RMS error of the estimated respiratory rate
$\tilde{B}(n)$ compared to the ground truth, (ii) the percentage of estimated
$\tilde{B}(n)$ within $3$ BPM of the ground truth, and (iii) the absolute
correlation of the estimated waveform $\tilde{r}(n)$ compared to the ground
truth. Comparisons are performed exclusively at the time steps $n$, where each
method yields a respiratory rate and/or waveform estimate. Of these
state-of-the-art methods, only Pos-Free Breath and Beamform estimate a
respiratory waveform. Therefore, respiratory waveform estimation is compared
only for Pos-Free Breath, Beamform, and WiRM. We use the absolute correlation
value because some of these methods may produce a waveform that is inverted to
the ground truth. In such cases, the correlation is negative but still indicates
a strong match.

In general, the results using existing methods are less favourable compared to
those reported in the literature because the latter is based on an indicator if
breath has been detected or not. The accuracy of existing methods has previously
been evaluated at time steps where the method has detected that a breath is
present. However, to maintain fairness to all methods (given that the datasets
used \textit{always} contain breathing), we evaluate the accuracy for all
timestamps, not just those where a breath is detected. Additionally, the
Pos-Free Breath algorithm was originally evaluated on a CSI dataset containing
the two receive antennas at two different perpendicular locations relative to
the chest. In this work we evaluate it on a standard antenna configuration.

\subsection{Parameter Settings}
\label{ssec:param_settings}
The parameters specific to our proposed algorithm are configured as follows. The
amount of samples collected for respiratory rate estimation $N = 150$, the
number of windows $W = 46$, the overlap of each window $Y = 140$ (i.e.,
approximately $1$ second between respiration rate estimates), the AMTC algorithm
smoothing parameter $\Lambda = 0.5$, the number of samples collected for
respiratory waveform estimation $\tilde{N} = 600$, and the fast iterative
filtering tuning parameter $\chi = 3.8$. The maximum and minimum respiratory
rates are defined as $[f_\text{b,min}, f_\text{b,max}] = [0.133\text{Hz},
0.833\text{Hz}]$, which corresponds to approximately $8\text{BPM}$ and
$50\text{BPM}$ respectively.

The parameters that are common amongst all methods are derived from the
open-source dataset outlined in Subsection~\ref{sssec:rf_poly_dataset}. These
common parameters are listed in Table~\ref{tab:common_parameter_settings}.

\begin{table}[tb!]
    \vspace{\abovetable}
    \caption{Common Parameter Settings}
    \label{tab:common_parameter_settings}
    \vspace{\betweentablecaption}
    \centering
    \begin{tabular}{@{} l l @{}}
        \toprule
        Parameter Name & Value \\
        \midrule
        Number of subcarriers ($K$) & 114 \\ 
        Sample frequency ($f_s$) & $9.9\text{Hz}$\\
        Number of  transmit antennas ($A$) & 2 \\
        Number of  receive antennas ($B$) & 2 \\
        Number of  breaths & 1 \\
        \bottomrule
    \end{tabular}
    \vspace{\belowtable}
\end{table}

\subsection{Respiratory Rate Performance}
\subsubsection{Dataset}
\label{sssec:rf_poly_dataset}
The public and open-source radio frequency polysomnography dataset
\cite{ref:csi_poly_dataset} is used to evaluate the performance of each method.
It was created for contactless respiration monitoring using radio frequency
measurements. The dataset consists of 20 individuals having a polysomnography
examination, while the CSI and other radio frequency parameters are recorded
simultaneously. Each individual is only monitored for a single sleeping period,
giving a total of 20 nights of data. We use ten nights of data for our
analysis, referring to each night of data as a test subject.

The estimation results calculated from this dataset may contain notable
variation across test subjects. This can be explained by the natural variation
of breathing waveforms and sleeping positions of different people. This can lead
to significant variation in respiratory rate and waveform estimation, especially
if some patients breathe more shallowly, or move frequently throughout the night.

The polysomnography data consist of measurements from a chest rip, abdomen rip,
thermocouple, and nasal cannula. The ground truth respiratory waveform is the
direct chest rip measurement, whereas the respiration rate is derived from the
polysomnography measurements using both low-pass filtering and the zero-crossing
method.

The CSI data consist of complex numbers representing the gain and phase shift
of the channel at the current time step and subcarrier, as measured by the NIC.
These data are used directly as the input to each respiration monitoring algorithm
with no pre-processing.

\subsubsection{Results}
\label{sssec:rms_results}

The RMS error of each method for each of the ten test subjects of radio
frequency polysomnography data is shown in Figure~\ref{fig:rms_per_day}. It can
be seen that WiRM outperforms all other methods, regularly achieving a lower RMS
error per test subject. When compared against the best-performing state-of-the-art
method, WiRM achieves an overall average RMS error improvement of $34.7\%$.

\begin{figure}
    \centering
    \includesvg[width=\columnwidth]{./Respiratory_Rate_RMS_Error_Per_Test_Subject}
    \vspace{\belowfig}
    \caption{
        The RMS error measured in BPM of the estimated respiratory rate
        $\tilde{B}(n)$ versus the ground truth radio frequency polysomnography
        data $B(n)$ for ten test subjects.
    }
    \vspace{\belowfigcaption}
    \label{fig:rms_per_day}
\end{figure}

\begin{figure}
    \centering
    \includesvg[width=\columnwidth]{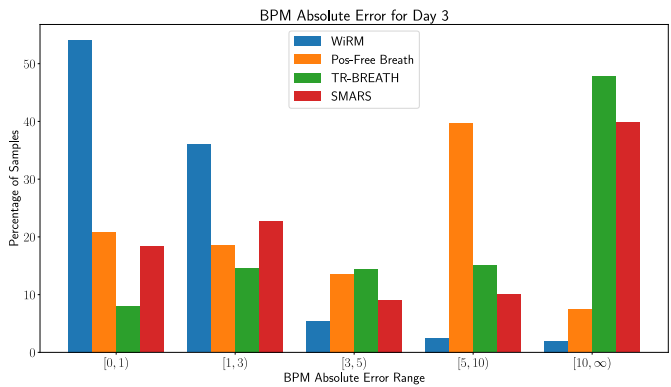}
    \vspace{\belowfig}
    \caption{
        Percentage of samples where each method achieves an absolute error
        within specified ranges for one test subject. The vertical axis shows
        the proportion of samples relative to the total sample count.
    }
    \vspace{\belowfigcaption}
    \label{fig:our_bpm_error_bins}
\end{figure}

Another useful performance metric is the percentage of estimations within a
given absolute error of the ground truth. An example of this metric for each
method is presented in Figure~\ref{fig:our_bpm_error_bins}. As shown, for the
provided test subject, WiRM has over $50\%$ of its samples within $1\text{ BPM}$
of the ground truth respiratory rate and over $30\%$ within the error range of
$[1, 3)\text{ BPM}$. These results show that WiRM closely tracks the
ground-truth respiratory rate for most estimates from the test subject.

This ability for WiRM to track the respiratory rate over time is further
displayed in Figure~\ref{fig:our_bpm_estimations}, where the estimated
respiratory rate $\tilde{B}(n)$ is closely following the ground truth $B(n)$.
Figure~\ref{fig:our_bpm_estimations} also contains red shading where the method
was unable to detect the presence of breathing. Notably, WiRM correctly
identifies the presence of a breath within the CSI data for $93.0\%$ of the time
for the given test subject.

\begin{figure}
    \centering
    \includesvg[width=\columnwidth]{./Respiratory_Rate_vs_Estimated_Rate}
    \vspace{\belowfig}
    \caption{
        An example of WiRM tracking the ground truth respiratory rate $B(n)$ for
        a single test subject from the radio frequency polysomnography dataset.
    }
    \vspace{\belowfigcaption}
    \label{fig:our_bpm_estimations}
\end{figure}

\begin{figure}
    \centering
    \includesvg[width=\columnwidth]{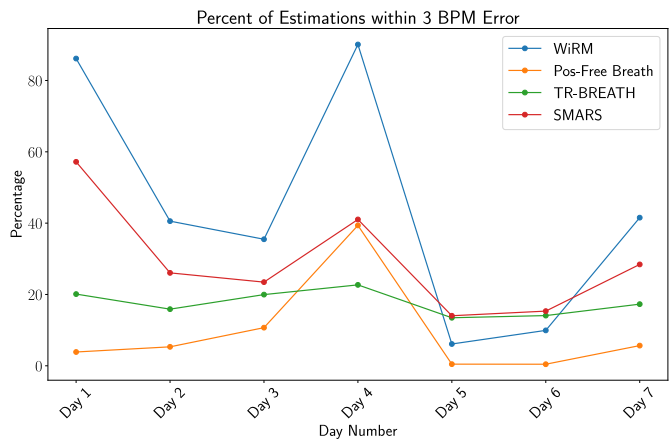}
    \vspace{\belowfig}
    \caption{
        The percentage of time that each method is within an absolute
        error of $3$ BPM compared to the ground truth of $B(n)$.
    }
    \vspace{\belowfigcaption}
    \label{fig:estimates_within_bpm}
\end{figure}

Performing a similar analysis for all ten test subjects, measuring the
percentage of samples within $3\text{ BPM}$ of the ground truth respiratory rate
$B(n)$ yields Figure~\ref{fig:estimates_within_bpm}. For most of the test
subjects from the radio-frequency polysomnography dataset, WiRM is observed to
have a higher percentage of samples within $3\text{ BPM}$ of absolute value
error. Although not outperforming the current state-of-the-art for all of the
test subjects against this metric, it still achieves an average improvement of
$30.4\%$ across all ten test subjects when compared to the best performing
method for each test subject.

\subsection{Respiratory Waveform Performance}
\subsubsection{Dataset}
The performance evaluation of the respiratory waveform is conducted using the
same open-source dataset described in Subsection~\ref{sssec:rf_poly_dataset}.

\subsubsection{Results}
\label{sssec:corr_results}
Here we compare the average absolute correlation $|\rho|$ of the produced
waveform $\tilde{r}(n)$ with the ground truth waveform $r(n)$ obtained from the
chest rip of the radio frequency polysomnography dataset. The average absolute
correlation is computed as the mean of $|\rho_{r\tilde{r}}(n)|$ from
\eqref{eq:abs_correlation}, for each test subject, averaged over a sleeping
period of approximately 11 hours. The values of $|\rho|$ are shown in
Figure~\ref{fig:absolute_corr_per_day}, where the maximum achieved correlation
per test subject is also displayed. It is evident from the figure that WiRM
improves the average and maximum correlation by $7.9\%$ and $1.9\%$,
respectively, compared to the state-of-the-art approach.

\begin{figure}
    \centering
    \includesvg[width=\columnwidth]{./Respiratory_Waveform_Absolute_Correlation_Per_Test_Subject}
    \vspace{\belowfig}
    \caption{
        The average and maximum absolute correlation values per test subject
        comparing WiRM, Pos-Free Breath, and Beamform.
    }
    \vspace{\belowfigcaption}
    \label{fig:absolute_corr_per_day}
\end{figure}

\subsection{Robustness Against Noise}
\subsubsection{Simulation Toolkit and Simulated Dataset}
\label{sssec:simulated_data}
In this paper, we develop a simulation toolkit for fast and consistent
comparison of wireless respiration-monitoring methods under various forms of
noise and ambient motion interference. Based on
\eqref{eq:non_ideal_system_model}, it allows configurable control over noise,
static and dynamic components, including the breath waveform and the number of
breaths. All stochastic elements are seeded for reproducibility. Simulation
configurations are stored with results to ensure consistent conditions across
different estimation algorithms.

The simulated data are idealistic in their construction, allowing for scenarios
where the error of some of the estimation methods is zero. This is neither
indicative of the performance in the real world nor the purpose of the
simulation. Rather, the simulation allows for the behaviour of different
respiration monitoring solutions to be compared when subjected to ambient
motion interference and three kinds of noise: thermal (additive),
multiplicative, and phase. Since the ability to vary noise precisely and measure
its impact is not currently available in the literature, it proves valuable in
evaluating the robustness of respiration monitoring methods. The performance of
WiRM with respect to non-ideal noise scenarios can be inferred from the results
against the real-world dataset in Subsections~\ref{sssec:rms_results} and
\ref{sssec:corr_results}. The simulated data allows us to compare different
estimation algorithms against different types of noise in isolation.

Each type of noise is generated as described in
Subsection~\ref{sec:system_model}. The noise levels can be varied by performing
simulations where the only changing parameter is the standard deviation of the
noise or in the case of phase noise, whether it is present or not.

Interference from ambient motion is evaluated by injecting an additional
multipath component into the simulated CSI in \eqref{eq:system_model}. The resulting
CSI is given by 
\begin{equation}
    \begin{aligned}
        \label{eq:ambient_motion_model}
        \hat{H}_{T_a, R_b}(t, f) = H_{T_a, R_b}(t, f) + &\alpha'_{T_a, R_b}(t, f)\\ 
                                                        &\times
                                                        \exp\Big(\frac{-j 2 \pi}{\lambda} D'_{T_a, R_b}(t) \Big),
    \end{aligned}
\end{equation}
where ${\alpha'_{T_a, R_b}(t, f) \in \mathbb{R}^{+}}$ represents the time-varying
magnitude of the interference and $D'_{T_a, R_b}(t)$ represents the time-varying
propagation distance of the interfering multipath. 

The interference magnitude is modelled using the single-slope path loss
model, ${\alpha'_{T_a, R_b}(t, f) = \Gamma \alpha_{T_a,
R_b}(f)}D'(t)^{-\gamma/2}$, where $\Gamma \in \mathbb{R}^+$ scales the
interference relative to the breathing multipath magnitude $\alpha_{T_a,
R_b}(f)$, and $\gamma$ is the single-slope path loss exponent. We set
$\gamma$ to a value representative of an indoor office
environment~\cite{ref:andrea_goldsmith}, as listed in
Table~\ref{tab:noise_simulation_config}. To model a moving person, $D'_{T_a,
R_b}(t)$ varies as a triangle wave between $1$ m and $10$ m, representing a
person walking towards and away from the receiver. The period of the
triangle wave is chosen to match natural walking speeds
\cite{ref:walking_speeds}. The range of walking speeds is provided in
Table~\ref{tab:noise_simulation_config}.

The parameters for all simulations described in
Section~\ref{ssec:simulation_results} are chosen to align with the public open
source dataset as shown in Table~\ref{tab:common_parameter_settings}. An
ideal sinusoidal respiratory waveform is chosen as $r(n)$. We select the
remaining simulation parameters as shown in
Table~\ref{tab:noise_simulation_config}.

\begin{table}[tb!]
    \vspace{\abovetable}
    \caption{Simulation parameters}
    \label{tab:noise_simulation_config}
    \vspace{\betweentablecaption}
    \begin{tabular}{@{} l l @{}}
        \toprule
        Parameter Name & Value \\
        \midrule
        Noise tests time & Time for 10 estimations \\
        Interference tests time & 3 minutes \\
        Antenna separation & $0.05$m\\
        Transmitter-receiver separation & $5$m \\ 
        Number of breaths & 1 \\
        Complex amplitude ($\alpha_{T_a, R_b}(f)$)& $U([0.2, 2])^\dagger$ \\
        Breath amplitude ($\Delta d_{\text{b},T_a, R_b}$)& $U([0.005, 0.01])^\ddagger$ \\
        Breath frequency BPM ($B_\text{Hz}(n)$)& 15 BPM \\
        Breath phase offset & 0 \\
        Static component amplitude ($|H_{\text{s},T_a,R_b}|$)& $U([10, 20])$ \\
        Static component phase ($\angle H_{\text{s},T_a,R_b}$)& $U([0, 2\pi))$ \\
        Dynamic component phase & $U([0, 2\pi))$ \\
        Dynamic component rotation direction &  $U(\{-1,+1\})$ \\
        Path loss exponent ($\gamma$) & $U([1.6, 3.5))$ \\
        Interfering walk speed & $U([1.21\mathrm{ms}^{-1}, 1.34\mathrm{ms}^{-1})) $ \\
        \bottomrule
    \end{tabular}

    $^\dagger$ $U(\mathcal{M})$ denotes a uniform distribution over the set $\mathcal{M}$.\newline
    $^\ddagger$ These values correspond to phase angles close to $\pi/3$ following chest motion assumptions \cite{ref:human_resp_detection_with_commodity_wifi}.
    \vspace{\belowtable}
\end{table}

To produce the results in Subsection~\ref{ssec:simulation_results}, each
simulation is run fifteen times using different random seeds, generating new
environmental conditions. For the noise robustness evaluation, sufficient CSI
estimates are generated for each method to produce 10 respiratory estimates. For
interference robustness evaluation, three minutes of CSI estimates are
generated. For each simulation, we compute the respiratory rate RMS error and
the absolute correlation of the estimated respiratory waveform. Results are
averaged over fifteen random seeds. The process is repeated for different levels
of thermal, multiplicative, and phase noise, as well as different interference
scaling factors $\Gamma$.

\subsubsection{Results}
\label{ssec:simulation_results}

Using the simulated dataset, the average respiratory rate RMS errors are shown
in Tables \ref{tab:sim_rms_phase_noise}, \ref{tab:sim_rms_multiplicative_noise},
\ref{tab:sim_rms_thermal_noise} and \ref{tab:sim_rms_interference},
whereas the average absolute correlation of the respiratory waveform in Tables
\ref{tab:sim_corr_phase_noise}, \ref{tab:sim_corr_multiplicative_noise},
\ref{tab:sim_corr_thermal_noise}, and \ref{tab:sim_corr_interference}. All
results are reported as value (standard deviation). Tables
\ref{tab:sim_rms_phase_noise} and \ref{tab:sim_corr_phase_noise} show that Pos-Free
Breath, Beamform, SMARS, and WiRM are unaffected by phase noise.
This robustness arises either because phase information is not used, as in SMARS, or
because phase noise cancellation techniques are applied. In contrast, TR-BREATH
does not mitigate phase noise, resulting in degraded performance.

The performance of the WiRM method in the face of multiplicative noise is shown
in Tables \ref{tab:sim_rms_multiplicative_noise} and
\ref{tab:sim_corr_multiplicative_noise}. As seen, WiRM is particularly resilient
to this type of noise thanks to the use of both the magnitude and phase of the
CSI data. To provide an intuition for the resilience to multiplicative noise, we
consider a simplified scenario from
\eqref{eq:non_ideal_system_model} as
\begin{equation}
    \label{eq:only_multiplicative}
    \tilde{H}_{T_a, R_b}(t, f) = \kappa(t, f) H_{T_a, R_b}(t, f),
\end{equation}
where $\kappa(t, f) \in \mathbb{R}^+$ is the log-normally distributed
multiplicative noise. The magnitude of the CSI is given by $|\tilde{H}_{T_a,
R_b}(t, f)| = \kappa(t, f)|H_{T_a, R_b}(t, f)|$, leaving the phase unaffected as
${\angle\tilde{H}_{T_a, R_b}(t,f) = \angle H_{T_a, R_b}(t, f)}$.

Considering the Fourier transform of the magnitude, we have
$\mathcal{F}(|\tilde{H}_{T_a, R_b}(t, f)|) = \mathcal{F}(\kappa(t, f)) *
\mathcal{F}(|H_{T_a, R_b}(t, f)|)$, where $*$ is the convolution operator. It
can be seen that the spectrum $\mathcal{F}(|H_{T_a, R_b}(t, f)|)$ is mixed with
the spectrum $\mathcal{F}(\kappa(t, f))$. Given $\kappa(t, f)$ has a log-normal
distribution with a non-trivial Fourier transform, intuitively the energy within
the resulting spectrum will spread or smear across a wider range of frequencies.
Conversely, the Fourier transform of the phase will be unaffected by the
multiplicative noise. The same intuition can be applied to the conjugate
multiplied CSI. This intuition is used in WiRM to counteract multiplicative
noise. Due to the smearing encountered in the spectrum of the magnitude, the BNR
will be lower for the magnitude when compared to the unaffected phase. This
leads to the magnitude columns in \eqref{eq:bnr_combining} being weighted with a
smaller BNR compared to the phase columns, accentuating the noise-free phase and
diminishing the noise affected magnitude. The same intuition can be applied to
the subcarrier combining approach described in
Subsection~\ref{sssec:subcarrier_combining}. It is noteworthy that the other
methods do not have this property of considering magnitude and phase separately
and, as such, are greatly affected by multiplicative noise.

Tables \ref{tab:sim_rms_interference} and \ref{tab:sim_corr_interference}
show the performance of WiRM under environmental interference. WiRM
effectively suppresses interference from a simulated person walking in the
room and achieves competitive performance. However, its performance degrades
when the interference scaling factor $\Gamma$ exceeds $5$. It is noted that
Beamform is particularly robust to interference and maintains strong
respiratory waveform estimation performance.

We note that all of the considered methods have superior performance on the
simulated data than on the real-world dataset. While the simulated noise is
always uncorrelated, white, and ideal, this is not always the case in the real
world as the noise among antennas can potentially be correlated, or the noise
may not necessarily be white. Additionally, for the purpose of simulation, the
breath waveform is an ideal sinusoid with a fixed frequency of $15$ BPM and
there are no other movements in the simulated environment. This is unrealistic
in a practical scenario as people breathe with varying waveforms, and there are
often other movements in the room that may cause undue interference. Despite
these differences, the simulated results still provide a useful metric for
comparison of each method when examined under idealistic scenarios, allowing the
impacts of certain classes of noise to be understood.

WiRM maintains the same level of robustness to phase noise as existing methods
such as Pos-Free Breath while also retaining a similar level of robustness to
thermal noise. Notably, WiRM exceeds the current state-of-the-art in robustness
to multiplicative noise, as shown by the improved RMSE for the respiratory rate
estimates and the improved absolute correlation for the respiratory waveform. We
note that WiRM still has similar limitations as the other methods in terms of
robustness to ambient motion. All of these results indicate that WiRM is a
suitable algorithm for uses such as sleep studies. 

\begin{table}[tb!]
    \vspace{\abovetable}
    \centering
    \caption{Respiratory Rate RMS Error (BPM): Phase Noise}
    \vspace{\betweentablecaption}
    \label{tab:sim_rms_phase_noise}
    \begin{tabular}{@{} l l @{}}
        \toprule
        Method & RMS Error (BPM)\\
        \midrule
        WiRM & $0.04\ (0.0)$ \\
        Pos-Free Breath & $0.67\ (1.9)$ \\
        Beamform & $\mathbf{0.0}\ (0.0)$ \\
        TR-BREATH & $6.44\ (2.6)$ \\
        SMARS & $0.83\ (0.0)$ \\
        \bottomrule
    \end{tabular}
    \vspace{\belowtable}
\end{table}

\begin{table}[tb!]
    \vspace{\abovetable}
    \centering
    \caption{Respiratory Rate RMS Error (BPM): Multiplicative Noise}
    \label{tab:sim_rms_multiplicative_noise}
    \vspace{\betweentablecaption}
    \begin{tabular}{@{} L{0.5cm} L{1.2cm} L{1.2cm} L{1.4cm} L{1.2cm} L{1.5cm} @{}}
        \toprule
        \multirow{2}{*}{\parbox[c]{0.5cm}{\centering Std\newline Dev}} & \multicolumn{5}{c}{RMS Error (BPM)}\\
        \cmidrule{2-6}
        \phantom{abc} & WiRM & Pos-Free Breath & Beamform & TR-BREATH & SMARS\\
        \midrule
        $0.1$ & $0.04\ (0.0)$ & $\mathbf{0.03}\ (0.0)$ & $6.45\ (11.2)$ & $0.33\ (0.3)$ & $3.89\ (5.3)$\\
        $0.25$ & $\mathbf{0.04}\ (0.0)$ & $0.7\ (1.7)$ & $15.35\ (13.5)$ & $0.56\ (0.6)$ & $10.59\ (7.0)$\\
        $0.5$ & $\mathbf{0.04}\ (0.0)$ & $4.15\ (4.1)$ & $27.65\ (0.8)$ & $0.48\ (0.5)$ & $16.71\ (3.0)$\\
        $0.75$ & $\mathbf{0.04}\ (0.0)$ & $7.86\ (0.7)$ & $27.64\ (0.6)$ & $0.42\ (0.6)$ & $16.29\ (4.1)$\\
        $1.0$ & $\mathbf{0.04}\ (0.0)$ & $8.29\ (1.0)$ & $27.7\ (0.4)$ & $0.26\ (0.4)$ & $16.63\ (4.4)$\\
        \bottomrule
    \end{tabular}
    \vspace{\belowtable}
\end{table}

\begin{table}[tb!]
    \vspace{\abovetable}
    \centering
    \caption{Respiratory Rate RMS Error (BPM): Thermal Noise}
    \label{tab:sim_rms_thermal_noise}
    \vspace{\betweentablecaption}
    \begin{tabular}{@{} L{0.5cm} L{1.2cm} L{1.2cm} L{1.4cm} L{1.2cm} L{1.5cm} @{}}
        \toprule
        \multirow{2}{*}{\parbox[c]{0.5cm}{\centering Std\newline Dev}} & \multicolumn{5}{c}{RMS Error (BPM)}\\
        \cmidrule{2-6}
        \phantom{abc} & WiRM & Pos-Free Breath & Beamform & TR-BREATH & SMARS\\
        \midrule
        $0.1$ & $0.04\ (0.0)$ & $0.03\ (0.0)$ & $\mathbf{0.0}\ (0.0)$ & $0.52\ (0.6)$ & $0.91\ (0.1)$\\
        $0.5$ & $0.04\ (0.0)$ & $0.03\ (0.0)$ & $\mathbf{0.0}\ (0.0)$ & $1.23\ (1.5)$ & $0.92\ (0.1)$\\
        $1.0$ & $0.05\ (0.0)$ & $\mathbf{0.02}\ (0.0)$ & $1.92\ (7.2)$ & $2.41\ (2.7)$ & $3.44\ (4.6)$\\
        $5.0$ & $\mathbf{2.87}\ (3.4)$ & $7.34\ (0.7)$ & $23.52\ (9.5)$ & $6.95\ (3.0)$ & $14.06\ (5.5)$\\
        $10.0$ & $\mathbf{5.17}\ (2.9)$ & $7.72\ (0.6)$ & $27.33\ (0.9)$ & $7.25\ (2.2)$ & $18.08\ (3.7)$\\
        \bottomrule
    \end{tabular}
    \vspace{\belowtable}
\end{table}

\begin{table}[tb!]
    \vspace{\abovetable}
    \centering
    \caption{Respiratory Rate RMS Error (BPM): Interference}
    \label{tab:sim_rms_interference}
    \vspace{\betweentablecaption}
    \begin{tabular}{@{} L{0.4cm} L{1.4cm} L{1.2cm} L{1.4cm} L{1.2cm} L{1.5cm} @{}}
        \toprule
        \multirow{2}{*}{\parbox[c]{0.0cm}{$\Gamma$}} & \multicolumn{5}{c}{RMS Error (BPM)}\\
        \cmidrule{2-6}
        \phantom{abc} & WiRM & Pos-Free Breath & Beamform & TR-BREATH & SMARS\\
        \midrule
        $1.0$ & $0.04\ (0.0)$ & $0.06\ (0.0)$ & $\mathbf{0.0}\ (0.0)$ & $0.37\ (0.7)$ & $0.87\ (0.1)$\\
        $1.5$ & $0.04\ (0.0)$ & $0.06\ (0.0)$ & $\mathbf{0.0}\ (0.0)$ & $0.64\ (0.9)$ & $0.86\ (0.1)$\\
        $2.0$ & $0.04\ (0.0)$ & $0.07\ (0.0)$ & $\mathbf{0.01}\ (0.0)$ & $1.38\ (1.9)$ & $0.84\ (0.1)$\\
        $5.0$ & $2.56\ (8.9)$ & $\mathbf{0.05}\ (0.0)$ & $5.69\ (11.7)$ & $1.52\ (2.7)$ & $3.36\ (2.9)$\\
        $10.0$ & $9.05\ (12.1)$ & $3.97\ (4.3)$ & $16.04\ (21.2)$ & $\mathbf{2.66}\ (4.3)$ & $9.2\ (5.3)$\\

        \bottomrule
    \end{tabular}
    \vspace{\belowtable}
\end{table}

\begin{table}[tb!]
    \vspace{\abovetable}
    \centering
    \caption{Respiratory Waveform Correlation ($|\rho|$): Phase Noise}
    \vspace{\betweentablecaption}
    \label{tab:sim_corr_phase_noise}
    \begin{tabular}{@{} l l @{}}
        \toprule
        Method & Correlation ($|\rho|$) \\
        \midrule
        WiRM & $\mathbf{1.0}\ (0.0)$ \\
        Pos-Free Breath & $0.97\ (0.1)$ \\
        Beamform & $\mathbf{1.0}\ (0.0)$ \\
        \bottomrule
    \end{tabular}
    \vspace{\belowtable}
\end{table}

\begin{table}[tb!]
    \vspace{\abovetable}
    \centering
    \caption{Respiratory Waveform Correlation ($|\rho|$): Multiplicative Noise}
    \label{tab:sim_corr_multiplicative_noise}
    \vspace{\betweentablecaption}
    \begin{tabular}{@{} l l l l @{}}
        \toprule
        \multirow{2}{*}{\parbox[c]{0.5cm}{\centering Std\newline Dev}} & \multicolumn{3}{c}{Correlation ($|\rho|$)}\\
        \cmidrule{2-4}
        \phantom{abc} & WiRM & Pos-Free Breath & Beamform\\
        \midrule
        $0.1$ & $\mathbf{1.0}\ (0.0)$ & $0.99\ (0.0)$ & $0.9\ (0.1)$\\
        $0.25$ & $\mathbf{1.0}\ (0.0)$ & $0.9\ (0.3)$ & $0.66\ (0.3)$\\
        $0.5$ & $\mathbf{1.0}\ (0.0)$ & $0.45\ (0.5)$ & $0.34\ (0.2)$\\
        $0.75$ & $\mathbf{1.0}\ (0.0)$ & $0.01\ (0.0)$ & $0.2\ (0.1)$\\
        $1.0$ & $\mathbf{1.0}\ (0.0)$ & $0.01\ (0.0)$ & $0.17\ (0.0)$\\
        \bottomrule
    \end{tabular}
    \vspace{\belowtable}
\end{table}

\begin{table}[tb!]
    \vspace{\abovetable}
    \centering
    \caption{Respiratory Waveform Correlation ($|\rho|$): Thermal Noise}
    \label{tab:sim_corr_thermal_noise}
    \vspace{\betweentablecaption}
    \begin{tabular}{@{} l l l l @{}}
        \toprule
        \multirow{2}{*}{\parbox[c]{0.5cm}{\centering Std\newline Dev}} & \multicolumn{3}{c}{Correlation ($|\rho|$)}\\
        \cmidrule{2-4}
        \phantom{abc} & WiRM & Pos-Free Breath & Beamform\\
        \midrule
        $0.1$ & $\mathbf{1.0}\ (0.0)$ & $0.99\ (0.0)$ & $1.0\ (0.0)$\\
        $0.5$ & $\mathbf{1.0}\ (0.0)$ & $0.99\ (0.0)$ & $0.96\ (0.1)$\\
        $1.0$ & $\mathbf{0.99}\ (0.0)$ & $0.99\ (0.0)$ & $0.88\ (0.1)$\\
        $5.0$ & $\mathbf{0.54}\ (0.4)$ & $0.02\ (0.0)$ & $0.43\ (0.2)$\\
        $10.0$ & $\mathbf{0.2}\ (0.3)$ & $0.01\ (0.0)$ & $0.17\ (0.1)$\\
        \bottomrule
    \end{tabular}
    \vspace{\belowtable}
\end{table}

\begin{table}[tb!]
    \vspace{\abovetable}
    \centering
    \caption{Respiratory Waveform Correlation ($|\rho|$): Interference}
    \label{tab:sim_corr_interference}
    \vspace{\betweentablecaption}
    \begin{tabular}{@{} l l l l @{}}
        \toprule
        \multirow{2}{*}{\parbox[c]{0.5cm}{$\Gamma$}} & \multicolumn{3}{c}{Correlation ($|\rho|$)}\\
        \cmidrule{2-4}
        \phantom{abc} & WiRM & Pos-Free Breath & Beamform\\
        \midrule
        $1.0$ & $\mathbf{1.0}\ (0.0)$ & $0.98\ (0.0)$ & $1.0\ (0.0)$\\
        $1.5$ & $\mathbf{1.0}\ (0.0)$ & $0.98\ (0.0)$ & $0.99\ (0.0)$\\
        $2.0$ & $\mathbf{1.0}\ (0.0)$ & $0.98\ (0.0)$ & $0.99\ (0.0)$\\
        $5.0$ & $0.93\ (0.3)$ & $\mathbf{0.98}\ (0.0)$ & $0.92\ (0.1)$\\
        $10.0$ & $0.39\ (0.5)$ & $0.34\ (0.5)$ & $\mathbf{0.76}\ (0.2)$\\
        \bottomrule
    \end{tabular}
    \vspace{\belowtable}
\end{table}

\subsubsection{Ablation Study}
The following ablation study is performed to show the impact of using the
proposed Gaussian frequency transition probability defined in
\eqref{eq:breath_distribution}. We compare this to a uniform transition
probability, where all transitions are equally likely. The two transition
probability functions are compared in a simulated scenario with varying levels
of thermal noise. Only thermal noise is considered since the other types of
noise have minimal impact, as discussed in
Subsection~\ref{ssec:simulation_results}.

As can be seen from Table~\ref{tab:ablation_thermal_noise}, the proposed
Gaussian frequency transition probability from \eqref{eq:breath_distribution}
adds a noticeable improvement in the RMSE for respiratory rate estimation,
indicating that the proposed method is superior in tracking breath.

\begin{table}[tb!]
    \vspace{\abovetable}
    \centering
    \caption{
        Frequency Transition Probability Function Ablation Study: Thermal Noise
    }
    \label{tab:ablation_thermal_noise}
    \vspace{\betweentablecaption}
    \begin{tabular}{@{} l l l @{}}
        \toprule
        \multirow{2}{*}{\parbox[c]{0.5cm}{\centering Std\newline Dev}} & \multicolumn{2}{c}{RMS Error (BPM)}\\
        \cmidrule{2-3}
        \phantom{abc} & Gaussian Distribution & Uniform Distribution\\
        \midrule
        $0.1$ & $\mathbf{0.04}\ (0.0)$ & $0.05\ (0.0)$\\
        $0.5$ & $\mathbf{0.04}\ (0.0)$ & $0.1\ (0.1)$\\
        $1.0$ & $\mathbf{0.05}\ (0.0)$ & $0.13\ (0.1)$\\
        $5.0$ & $\mathbf{2.62}\ (3.2)$ & $6.42\ (8.3)$\\
        $10.0$ & $\mathbf{5.87}\ (3.3)$ & $8.19\ (5.1)$\\
        $100.0$ & $\mathbf{8.63}\ (7.0)$ & $12.47\ (8.1)$\\
        \bottomrule
    \end{tabular}
    \vspace{\belowtable}
\end{table}

\subsection{Real-Time Feasibility Analysis}
\label{ssec:real_time_feasibility}
Given the complexity analysis described in
Subsection~\ref{sssec:resp_rate_comp_complexity} and
Subsection~\ref{sssec:resp_waveform_comp_complexity}, it could be assumed
that the proposed algorithm incurs a high computational cost due to the
often quadratic or quasilinear complexity. Fortunately, the constants
present in the complexity analysis of WiRM are in reality quite small as
described in Subsection~\ref{ssec:param_settings}.

To analyse the real-time feasibility of WiRM we use the simulation framework
described in Subsection~\ref{sssec:simulated_data} to generate $10,000$
respiratory rate and respiratory waveform estimations with the WiRM algorithm.
The runtime of each estimation is measured and the mean and standard deviation
are reported as mean (standard
deviation). This is repeated for two CPU's, first a personal computer (PC) with
an AMD Ryzen 9 7900X CPU, and a laptop with an Intel i7-8850H CPU. The results
are presented in Table~\ref{tab:real_time_feasibility}.

The WiRM algorithm is configured to produce a respiratory rate and waveform
estimation once every second or every $10$ samples, as described in
Subsection~\ref{ssec:param_settings}. This means that the computation time
of each WiRM estimation is well within the one second time window, making
WiRM feasible for real-time applications.

\begin{table}[tb!]
    \vspace{\abovetable}
    \caption{WiRM Real-Time Performance}
    \label{tab:real_time_feasibility}
    \vspace{\betweentablecaption}
    \centering
    \begin{tabular}{@{} l l l @{}}
        \toprule
        Estimation & \multicolumn{2}{c}{Computation Time (ms)}\\
        \cmidrule{2-3}
        \phantom{abc} & PC & Laptop\\
        \midrule
        Respiratory Rate  & $32.48\ (0.4)$ & $68.80\ (1.5)$ \\
        Respiratory Waveform  & $70.18\ (37.1)$ & $153.73\ (89.4)$ \\
        \bottomrule
        \bottomrule
        Total Time & $102.66\ (37.18)$ & $222.53\ (89.44)$ \\ 
        \bottomrule
    \end{tabular}
    \vspace{\belowtable}
\end{table}

\subsection{Discussion}
\label{ssec:discussion}

The WiRM algorithm demonstrates a strong balance between utility and performance
when compared to the four state-of-the-art methods. Its utility lies in the
rapid computation of both respiratory rate and waveform, while its performance
is marked by robustness to noise and high accuracy. In this subsection, we
examine the strengths and limitations of WiRM relative to these methods. A brief
overview of each method is provided in Section~\ref{sec:related_works}.

Pos-Free Breath uses the CSI-ratio, PCA, and VMD to decompose a respiratory
waveform from measured CSI data. WiRM has built upon this decomposition idea by
realising a guided decomposition. Rather than decomposing the measurement data
and picking the IMF that has a high variance as the respiratory waveform, WiRM
employs prior knowledge of the respiratory rate to search for the waveform that
best matches the expected rate. The results show that this method can lead to a
more accurate decomposition of the respiratory waveform and improved performance
in respiratory rate estimation. Both Pos-Free Breath and WiRM use a phase noise
cancellation technique that requires multiple antennas attached to the same
local oscillator. This unfortunately requires sacrificing some of the sensing
links for phase noise cancellation.

Beamform uses the genetic algorithm to compute beamforming weights that
enhance the desired signal while cancelling phase noise. Unlike CSI-ratio
and conjugate multiplication techniques, Beamform mitigates phase noise
without introducing additional cross-components. This likely explains its
strong resistance to ambient motion interference
(Table~\ref{tab:sim_corr_interference}) and its competitive respiratory
waveform correlation performance (Figure~\ref{fig:absolute_corr_per_day}).
Despite not requiring the genetic algorithm to be applied to each subcarrier
for every estimation, WiRM achieves similar or better waveform
correlation and respiratory rate RMSE than Beamform.

The TR-BREATH algorithm is among the first algorithms to realise multi-person
respiratory rate estimation. Although it is only capable of estimating the
respiratory rate (but not the respiratory waveform), the ability to
simultaneously estimate the rates of multiple people is an advantage over WiRM.
The TR-BREATH algorithm is more time consuming, with its use of the Root-MUSIC
algorithm and its cubic time complexity. TR-BREATH remains useful for offline
respiration estimation, whereas WiRM is better suited for real-time monitoring.

The SMARS algorithm is the fastest of the five methods evaluated in this
work. Although limited in its output with only respiratory rate estimation, its
simplicity in algorithm design makes it efficient for real-time estimation. The
majority of the algorithm is in applying the autocorrelation function and
peak-finding algorithms. Compared to WiRM which must sacrifice sensing links
between receive antennas for phase noise cancellation, SMARS is capable of
running on just a single link thanks to its use of only the CSI magnitude.
Unfortunately, this also introduces blind-spots in its estimation capabilities
\cite{ref:fullbreathe}. These blind-spots are resolved within the WiRM
algorithm, thanks to the use of both magnitude and phase simultaneously.

Wi-Fi-based methods for respiration monitoring can face challenges due to the
relatively low bandwidth of Wi-Fi. This can lead to poor spatial resolution,
requiring more complex algorithms to resolve multipath components reflecting off
different moving objects in the environment. For example, when monitoring the
respiration of multiple patients, the ambient motion from a busy or complex
environment may interfere with the received respiratory signal. Future work
should focus on addressing these problems to further the adoption of Wi-Fi-based
sensing methods.

\section{Conclusion} \label{sec:conclusion}
This paper has proposed a new method for the wireless monitoring of respiratory
health using Wi-Fi CSI. Our solution has improved upon the accuracy of the
current state-of-the-art respiratory rate estimation. We have demonstrated the
effectiveness of applying a respiratory rate estimate to improve the performance
of respiratory waveform estimation. We have also developed a simulation
framework that enables a fair comparison of various respiration monitoring
methodologies when subjected to different noise conditions.

\bibliographystyle{IEEEtran}
\bibliography{refs}

\vspace{-8mm}
\begin{IEEEbiographynophoto}{James Rhodes}
received the bachelor's degree in Electrical and Electronic Engineering (First
Class Honours) from The University of Newcastle, Australia, in 2022. He is
currently pursuing the Ph.D. degree with the School of Engineering, The
University of Newcastle. His research interests include digital signal
processing and integrated sensing and communications, particularly the use of
Wi-Fi as a contactless and widely available sensor for healthcare monitoring
applications.
\end{IEEEbiographynophoto}

\vspace{-8mm}
\begin{IEEEbiographynophoto}{Lawrence Ong}
(Senior Member, IEEE) received the BEng (First Class Honours) degree in electrical engineering from the National University of Singapore in 2001, the MPhil degree from the University of Cambridge in 2004, and the PhD degree from the National University of Singapore in 2008. He is currently an associate professor in the School of Engineering at the University of Newcastle, Australia. His research interests include information theory, secure communications, data privacy, and index coding. He was awarded a Discovery Early Career Researcher Award and a Future Fellowship from the Australian Research Council in 2012 and 2014, respectively. He served as an associate editor for the IEEE Transactions on Communications from 2018 to 2023.
\end{IEEEbiographynophoto}

\vspace{-8mm}
\begin{IEEEbiographynophoto}
{Duy Trong Ngo} (S'08-M'15-SM’20) received the B.Eng. (with First-class Honours and the University Medal) degree in telecommunication engineering from The University of New South Wales in 2007, the M.Sc. degree in electrical engineering (communication) from the University of Alberta in 2009, and the Ph.D. degree in electrical engineering from McGill University in 2013. He is currently an Associate Professor with the School of Engineering, The University of Newcastle, Australia. His research interests include wireless communications and machine learning applications.
\end{IEEEbiographynophoto}

\end{document}